\documentclass[proof]{pasj00}

\draft
\usepackage{times}
\usepackage{color}

\begin{document}
\SetRunningHead{Author(s) in page-head}{Running Head}

\title{Z45: A New 45-GHz Band Dual-Polarization HEMT Receiver
 for the NRO 45-m Radio Telescope}

%
\author{%
   Fumitaka \textsc{Nakamura}\altaffilmark{1,2}, 
   Hideo \textsc{Ogawa}\altaffilmark{3}, 
   Yoshinori \textsc{Yonekura}\altaffilmark{4}, 
   Kimihiko \textsc{Kimura}\altaffilmark{3},
   Nozomi \textsc{Okada}\altaffilmark{3},
   Minato \textsc{Kozu}\altaffilmark{3},
   Yutaka \textsc{Hasegawa}\altaffilmark{3},
   Kazuki \textsc{Tokuda}\altaffilmark{3},
   Tetsu \textsc{Ochiai}\altaffilmark{5},
   Izumi \textsc{Mizuno}\altaffilmark{6,7},
   Kazuhito \textsc{Dobashi}\altaffilmark{5},
   Tomomi \textsc{Shimoikura}\altaffilmark{5},
   Seiji \textsc{Kameno}\altaffilmark{8},
   Kotomi \textsc{Taniguchi}\altaffilmark{2,9}, 
   Hiroko \textsc{Shinnaga}\altaffilmark{1,7},
   Shuro \textsc{Takano}\altaffilmark{2,6,10}
   Ryohei \textsc{Kawabe}\altaffilmark{1},
   Taku \textsc{Nakajima}\altaffilmark{11}, 
   Daisuke \textsc{Iono}\altaffilmark{1},
   Nario \textsc{Kuno}\altaffilmark{12},
   Toshikazu \textsc{Onishi}\altaffilmark{3},
   Munetake \textsc{Momose}\altaffilmark{13},
   and
   Satoshi \textsc{Yamamoto}\altaffilmark{13}}
 \altaffiltext{1}{National Astronomical Observatory of Japan, 2-21-1 Osawa, Mitaka, Tokyo 181-8588}
\altaffiltext{2}{The Graduate University for Advanced Studies
(SOKENDAI), 2-21-1 Osawa, Mitaka, Tokyo 181-0015, Japan}
 \altaffiltext{3}{Department of Physical Science, Graduate School of
   Science, Osaka Prefecture University, 1-1 Gakuen-cho, Naka-ku, Sakai,
   Osaka}
\altaffiltext{4}{Center for Astronomy, Ibaraki University,
2-1-1 Bunkyo, Mito, Ibaraki 310-8512, Japan}
 \altaffiltext{5}{Department of Astronomy and Earth Sciences, 
Tokyo Gakugei University, 4-1-1 Nukuikitamachi, Koganei, Tokyo 184-8501}
   \altaffiltext{6}{Nobeyama Radio Observatory, National Astronomical Observatory of Japan
462-2 Nobeyama, Minamimaki, Minamisaku, Nagano 384-1305}
\altaffiltext{7}{Department of Physics, Faculty of Science, 
Kagoshima University, 1-21-35 Korimoto, Kagoshima, Kagoshima 890-0065, Japan}
\altaffiltext{8}{Joint ALMA Observatory, Alonso de C\'ordova 3107 Vitacura, Santiago, Chile}
 \altaffiltext{9}{Toho University, 2-2-1 Miyama, Funabashi, Chiba 274-8510, Japan}
 \altaffiltext{10}{Department of Physics, General Education, College of Engineering, Nihon
University, Tamuramachi, Koriyama, Fukushima 963-8642, Japan}
\altaffiltext{11}{Solar-Terrestrial Environment Laboratory, Nagoya University, Furo-cho,
Chikusa-ku, Nagoya, Aichi 464-8601, Japan}
\altaffiltext{12}{Department of Physics, Graduate School of Pure and Applied Sciences, 
The University of Tsukuba, 1-1-1 Tennodai, Tsukuba Ibaraki 305-8577,
   Japan}
\altaffiltext{13}{Institute of Astrophysics and Planetary Sciences, 
Ibaraki University, Bunkyo 2-1-1, Mito 310-8512, Japan}
\altaffiltext{13}{Department of Physics, Graduate School of Science, 
The University of Tokyo, Tokyo 113-0033, Japan}
 \email{fumitaka.nakamura@nao.ac.jp}


\maketitle

\begin{abstract}
We developed a dual-linear-polarization HEMT 
(High Electron Mobility Transistor) amplifier receiver
 system of the 45-GHz band (hereafter Z45),
and installed it in the Nobeyama 45-m radio telescope. 
The receiver system is designed to conduct polarization
 observations by taking the cross correlation of two linearly-polarized components, 
from which we process full-Stokes spectroscopy.
We aim to measure the magnetic field strength through the Zeeman effect of 
the emission line of CCS ($J_N=4_3-3_2$) toward pre-protostellar cores.
A linear-polarization receiver system has a smaller
contribution of instrumental polarization components to the Stokes
$V$ spectra than that of the circular polarization system, so that 
it is easier to obtain the Stokes $V$ spectra.
The receiver has an RF frequency of 42 $-$ 46 GHz and an intermediate frequency (IF)
 band of 4$-$8 GHz. The typical noise temperature is about 50 K, and 
the system noise temperature ranges from 100 K to 150K  over 
the frequency of 42 $-$ 46 GHz. 
The receiver system is connected to two spectrometers, SAM45 and PolariS.
SAM45 is a highly flexible FX-type digital spectrometer with a 
finest frequency resolution of 3.81 kHz.
PolariS is a newly-developed digital spectrometer with a finest 
frequency resolution of 60 Hz, having a capability to process
the full-Stokes spectroscopy.
The Half Power Beam Width (HPBW) of the beam was measured to be
 37$''$ at 43 GHz.
The main beam efficiency of the Gaussian main beam was derived to
 be 0.72  at 43 GHz. The SiO maser 
observations show that the beam pattern is reasonably round at about 10 \%
 of the peak intensity and the side-lobe level was less than 3 \%
 of the peak intensity.
Finally, we present some examples of astronomical observations using Z45.
\end{abstract}

\section{Introduction}
\label{sec:intro}

The NRO 45-m radio telescope is one of the world largest millimeter-wave 
single-dish telescopes 
operated by the Nobeyama Radio Observatory
(NRO) and located near the Yatsugatake mountains on the border of
Nagano Prefecture and Yamanashi Prefecture in Japan.
The observatory's longitude, latitude, and altitude are
138$^\circ$ 28$'$ 13$"$ (E), 35$^\circ$ 56$'$ 27$"$ (N), 
and 1349-m, respectively.
The telescope is equipped with various receivers for observations at
frequencies ranging from 20 GHz to 116 GHz, and used for various
research fields such as star and planet formation, astrochemistry, and 
evolution of galaxies.
However, no receivers that can conduct polarization observations are
available at present.
To open a new science window of polarization observations at millimeter
wavelengths using the NRO 45-m telescope, we developed a 40 GHz ($\approx 7$ mm) band,
dual-polarization receiver.
The dual-polarization receiver also has an advantage even for standard
molecular line observations; by observing two orthogonal polarized
components simultaneously, the sigal-to-noise ratio can be improved by a factor of
$\sqrt{2}$ with the same observation time. 

Magnetic fields are thought to play an important role in star formation (Shu et al. 1987).
However, the role of magnetic field in star formation process remains controversial (Crutcher et al. 2010; Crutcher 2012).
This is in part due to the difficulty of measuring the magnetic field strengths in molecular clouds.
Previous Zeeman observations toward star-forming regions have been conducted mainly at low frequencies using H$\,${\sc i} 
and OH lines (e.g., Crutcher et al. 1993; Heiles \& Troland 2004). These observations aim to measure the strengths of 
line-of-sight magnetic fields associated with interstellar clouds from the Stokes $V$ spectra. 
Strong lines like OH ($\approx 1.6$ GHz) and H$_2$O masers ($\approx 22$ GHz) have also been used for the Zeeman 
observations (e.g., Fiebig \& G\"usten 1989). These maser lines are thought to arise from the 
compact dense spots created by shocks near protostars. However, very few observations on molecules 
that probe dense molecular gas have been carried out so far (Crutcher et al. 1999; Falgarone et al. 2008). 
This is because among the molecules that have rotational transition lines with critical densities 
of 10$^4$ cm$^{-3}$, there are only a few molecules that have large magnetic dipole 
moments to show observable Zeeman splitting, e.g., CN, CCS, and SO. 
Crutcher et al. (1999) measured the Zeeman effect of the CN ($N=1-0$) lines at 113 GHz.
CN lines have critical densities of $10^5$ cm$^{-3}$ and thus is the suitable tracer to detect dense gas.  
The Zeeman observations at higher frequencies like 100 GHz are challenging because 
the Zeeman splitting is often smeared by the line broadening due to
internal motions. Therefore, the main targets of the CN Zeeman observations
are massive star-forming regions which sometimes emit intense emission lines of CN
and are expected to have stronger magnetic fields.
These observations provide us important information of the magnetic fields in star-forming regions.
However, the role of magnetic fields in the gravitational collapse of prestellar cores remain uncertain.
To measure the strengths of magnetic fields associated with dense molecular cloud cores
prior to star formation, we need other molecules that can trace the dense molecular gas 
and have relatively large dipole moments.  One of such molecules is thioxoethenylidene radical, CCS, 
which is a carbon-chain molecule and known to be abundant in molecular clouds before 
star formation happens (Suzuki et al. 1992). CCS has a critical density of $10^4$ cm$^{-3}$ and 
the Zeeman splitting is expected to be as large as 0.64 Hz/$\mu$G (Shinnaga \& Yamamoto 2000), because 
CCS has an unpaired electron.
CCS are often used as a good tracer of pre-protostellar cores and
CCS emission tends to be strongest at the 40 GHz band in molecular clouds  (Suzuki et al. 1992).

The 45 GHz band receivers designed to conduct the Zeeman observations are rare in the world.
The Zeeman observations at the 40 GHz band can be conducted with the Very Large Array (VLA, Momjian \& Sarma 2012).
Since VLA is an interferometer, the spatially-extended emission is significantly filtered out.
CCS emission is sometimes spatially extended, and therefore we consider that single dish telescopes have
an advantage to detect CCS Zeeman splitting toward dense molecular cloud cores. 
The 45-GHz receiver is also useful to conduct other observations.
For example, a number of carbon-chain molecules such as HC$_3$N, HC$_5$N and so on 
can often show strong emission at the 40 GHz band (Suzuki et al. 1992; Hirota \& Yamamoto 2006). These molecules are abundant 
at early evolutionary stages of molecular clouds before star formation is initiated.  
Therefore, their spatial distributions give us a hint to uncover the evolutionary stages of molecular clouds and cores.
Mapping observations using single dish telescopes are expected to be ideal to reveal the spatial distribution of
prestellar gas in molecular clouds.
Therefore, we consider it important to develop the new receiver at the 40 GHz band for the NRO 45-m telescope
to do new astronomical sciences.

In the present paper, we describe the basic characteristics 
of the newly-developed 40-GHz receiver system Z45
and show its performance. 
The detail and performance of the polarization observations
is presented in a forthcoming paper
of this project. Then, we describe the detail of the receiver system
in Section \ref{sec:develop}.
Observational results of the SiO masers from an evolved star, R-Leo, and continuum measurements
of Saturn are presented  in Section \ref{sec:results} and 
the size and shape of the system's primary beam and the main beam efficiency of the antenna are derived.
Then, we present in Section \ref{sec:test} some examples of 
observations of some molecular lines at the 40-GHz band.
These observations show the high performance of Z45.
Finally, we summarize the conclusion in Section \ref{sec:con}.

\section{A Primary Science Goal and Requirements}
\label{sec:science}

A primary science goal for the new receiver is to conduct the Zeeman
measurements toward dense cores prior to star formation, i.e.,
pre-protostellar cores or prestellar cores in short.

\subsection{Role of Magnetic Field in Gravitational Collapse of Prestellar Cores}

The dynamical stability of a magnetized prestellar core is determined by an
important parameter, mass-to-magnetic-flux ratio, $\mu \equiv M/ \Phi $ 
(e.g., Shu et al. 1987).
If the mass-to-flux ratio is larger than the critical value
$\mu_{\rm cr} \equiv 1/(2\pi G^{1/2}$), the magnetic field cannot support the 
whole cloud against self-gravity. Such a cloud is called
magnetically-supercritical and collapses dynamically to form  stars.
If the mass-to-flux ratio is smaller than the critical value,
the magnetic field can support the cloud against self-gravity.
Such a cloud is called magnetically-subcritical.
A magnetically-subcritical cloud can contract quasi-statically in losing 
the magnetic flux due to ambipolar diffusion. When the central dense 
part of the subcritical core becomes magnetically-supercritical, 
it collapses dynamically to form stars (e.g., Nakamura \& Li 2003).
Thus, the evolution of the magnetized cores is controlled by magnetic
field strength.
In this project, we aim to measure the strengths of the magnetic field associated with
prestellar cores through the Zeeman measurements and
constrain the role of magnetic field in the dynamical evolution of the cores.

\subsection{Zeeman Measurements with CCS}

The Zeeman effect can be observed as  the splitting of a spectral line into
several components in the presence of a magnetic field.
Since the displacement between the separated components can be given as
a function of the magnetic field strength, the Zeeman observations  allow
us to measure the magnetic field strength directly.
However, the detection of the Zeeman splitting toward prestellar
cores is challenging because the interstellar magnetic field is subtle and the splitting
frequency interval is narrow, often obscured by the
Doppler broadening due to internal motions of the cores. 
To detect the Zeeman splitting, we need spectral lines that have 
large magnetic dipole moments. The low frequency
lines have advantage to distinguish between the Zeeman splitting and the Doppler broadening.
Hence, the previous Zeeman measurements have been carried out mainly
with low-frequency H{$\,${\sc i}} and OH lines ($\approx 1-2$ GHz), or the strong maser lines. However,
the H{$\,${\sc i}} and OH lines can trace only low density gas of 
$\lesssim 10^2-10^3$ cm$^{-3}$. 
On the other hand, the maser lines can trace only the high density, 
high temperature spots near the stars. In
other words, our knowledge of the magnetic fields associated with dense molecular gas
remains limited. 
Recently, Falgarone et al. (2008) have successfully detected CN Zeeman
splitting at a 100 GHz band. CN is one of the dense molecular gas tracers with a
high critical density of $\approx 10^5$ cm$^{-3}$. 
The target clouds are, however, strongly biased toward
massive-star-forming regions or protostellar cores that have strong CN emission.
To understand the role of magnetic fields in initiating the
gravitational collapse of the cores, it is important to measure the
magnetic field of the cores that do not contain any stars yet,
i.e., prestellar cores.

One of the good tracers of high-density molecular gas in prestellar phase is 
thioxoethenylidene radical, CCS, which is
abundant in the early phase of core evolution prior to the protostellar formation. CCS
also has a high Zeeman splitting factor of its GHz transitions ($\approx
0.629$ Hz $\mu$G$^{-1}$ for $J_N=4_3-3_2$ transition, Shinnaga \& Yamamoto 2000).
In particular, the 45 GHz line is one of the strongest CCS lines in
prestellar cores with temperatures of about 10 K.
Even if the emission of $J_N=4_3-3_2$ (45 GHz) is relatively strong 
($\gtrsim 1$ K), the emission in the 100 GHz band (e.g., $J_N=8_7-7_6$) is often very weak ($\sim 0.1$ K) 
(see figure 1 of Wolkovitch et al. 1997). Although the 22.3
GHz CCS ($J_N = 2_1-1_0$) line sometimes has comparable strength to that of CCS ($J_N = 4_3-3_2$), 
the beam dilution effect is often significant because of the lower frequency. 
Therefore, the 45 GHz CCS line is likely to be most suitable for the Zeeman measurements.
Until now, Shinnaga et al. (1999b) reported the tentative detection of the magnetic field of 
160 $\pm$ 42 $\mu$G toward L1521E using the $J_N=4_3-3_2$ line ($\approx$ 45 GHz), and
Levin et al. (2001) reported 48 $\pm$ 31 $\mu$G toward L1498, using the $J_N=3_2-2_1$ 
line ($\approx$ 33 GHz).

To shed light on the issue of the role
of magnetic field in star formation, we have developed 
a new 40 GHz-band receiver system (hereafter Z45) for the NRO 45m
telescope and started a Zeeman measurement project, in which 
we attempt to measure the magnetic field strength using the Zeeman 
splitting of CCS ($J_N=4_3-3_2$).
It is worth noting that 
at a distance of nearby star forming regions like Taurus ($\approx$ 140
pc), the spatial resolution of the NRO 45-m telescope
at 45 GHz is 6000 AU ($\approx $ 0.03 pc), which is comparable to the
typical sizes of prestellar cores.

\begin{table}
  \caption{Prestellar cores having strong CCS emission}\label{tab:taurus}
  \begin{center}
    \begin{tabular}{lllllll}
     \hline
  Name & R.A. (J2000.0) & decl. (J2000.0) & $T_{\rm mb, peak}^*$ $^{**}$ (K) &
     $\Delta V$ (km s$^{-1}$) & $V_{\rm LSR}$ (km s$^{-1}$)& reference$^*$ \\
\hline
B1a  & 3:33:21 & 31:08:18& 1.6 & 0.23& 6.6 & 2  \\
B1b  & 3:33:21 & 31:08:48& 1.4 & 0.36& 6.6 & 2  \\
B1c  & 3:33:11.7 & 31:08:33& 1.8 & 0.21& 6.5 & 2  \\
L1495B & 4:15:42.4 & 28:47:48 & 3.6 & 0.30  & 7.6 & 2,5  \\
L1521B & 4:24:12.7& 26:36:53& 2.8& 0.40& 6.2 & 1,5 \\
L1521E & 4:26:12.5& 26:07:17& 2.5& 0.42& 6.6  &  3 \\
TMC-1 (NH$_3$) & 4:41:23.0& 25:48:13& 1.6& 0.64& 5.8 &1  \\
TMC-1 (CP) & 4:41:42.5& 25:41:27 & 3.2& 0.58& 5.9 & 1  \\
TMC-1C & 4:41:34.3 & 26:00:43 & 1.9 & 0.43   & 5.2 & 1  \\
L1544  & 5:04:15.3 & 25:11:48 & 1.6 & 0.41  & 7.2 & 1  \\
L492   & 18:15:46.1& -03:46:13 & 2.2 & 0.48 & 7.7 & 4  \\
CB130-3 & 18:16:17.9& -02:16:41 & 2.3& 0.48 & 7.2  & 6  \\
Serpens South-N1 &18:29:57.3 &-01:56:51 & 2.5& 0.9& 7.6 &7  \\
Serpens South-N2 & 18:29:57.1&-01:59:18& 2.8& 0.72& 7.4& 7  \\
Serpens South-C  & 18:30:01.3& -02:02:27& 1.6& 0.65& 7.0& 7 \\
Serpens South-S1 & 18:30:11.6& -02:06:14& 1.7& 0.42& 6.8 & 7 \\
Serpens South-S2 & 18:30:15.2& -02:07:45& 2.1& 0.81& 6.6 & 7 \\
L673-SMM4 & 19:20:24.6& 11:24:34 & 2.2& 0.48& 6.6 & 6 \\
\hline
\end{tabular}
\end{center}
$^*$ 1: Suzuki et al. (1992), 2: this study,
3: Hirota et al. (2002), 4: Hirota \& Yamamoto (2006),  5: Hirota et al. (2004), 6: Hirota et al. (2011),
 7: Nakamura et al. (2014)

$^{**}$These cores have peak intensity greater than 
$T_{\rm mb, peak}^* \ge $1.4 K, equivalent to $T_{\rm A, peak}^* \ge 1$ K for $\eta\sim 0.7$.
\end{table}

\begin{figure}
	   \begin{center}
  \includegraphics[height=10cm]{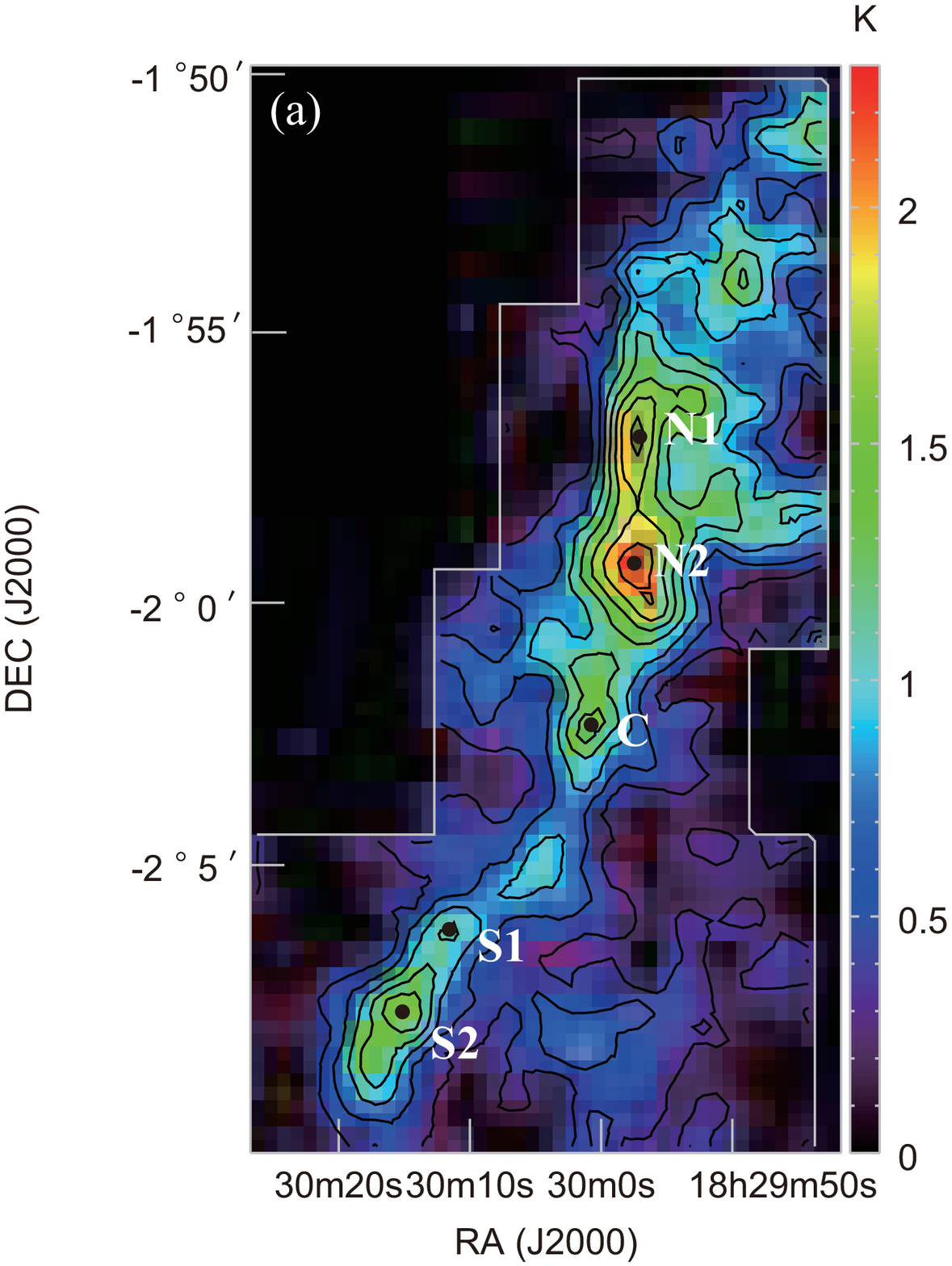} 
  \\
  \includegraphics[height=6cm]{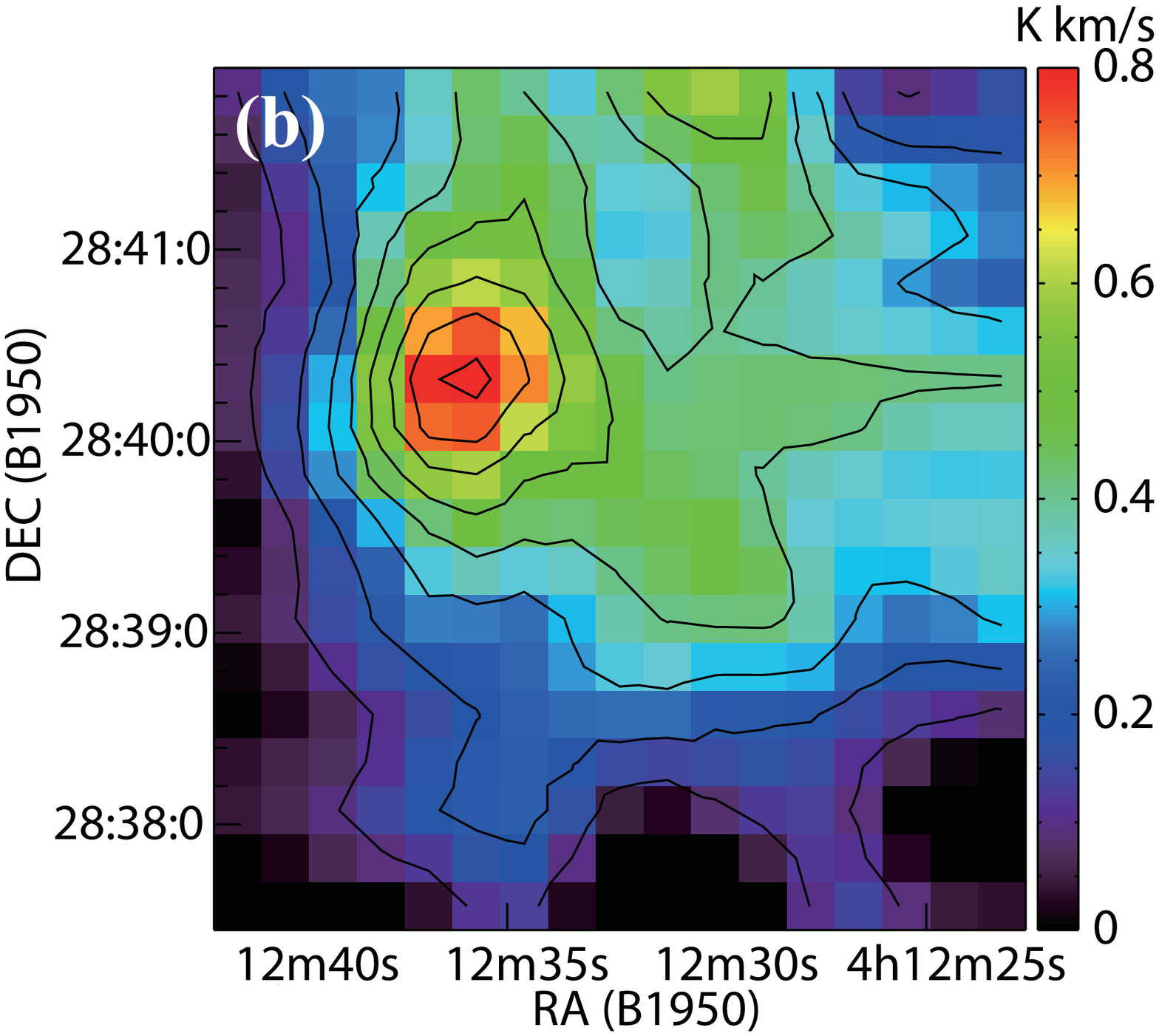} 
 \end{center}
\caption{(a) Integrated intensity map of CCS ($J_N=4_3-3_2$) toward Serpens South.
The contours start at 0.3 K km s$^{-1}$ with an
 interval of 0.3 K km s$^{-1}$. See Nakamura et al. (2014) for details.
(b) Integrated intensity map of CCS toward L1495B. The velocity range
 is 6 $-$ 9 km s$^{-1}$. The contours start from 0.1 K km s$^{-1}$ with an
 interval of 0.1 K km s$^{-1}$.}\label{fig:serps}
\end{figure}


We aim to measure the line-of-sight component of the magnetic fields using the Zeeman effect that appears in Stokes $V$ as 
\begin{equation}
V = \frac{dI}{d\nu} \Delta \nu . 
\end{equation}
Here $I$ is the Stokes $I$ profile, $\nu$ is the frequency, and $\Delta \nu$ is the Zeeman shift. Since the Stokes $V$ profile is proportional to the first derivative of the Stokes $I$, it is not significantly affected by contamination of Stokes $I$ through polarization leakage. Since Stokes $V$ is peaked at the both edges of Stokes $I$ profile, we need spectroscopy with a frequency resolution finer than the width of line edges (therefore finer than the full line width by ~2 orders of magnitude) to avoid spectral dilution.
Linear polarization (Stokes $Q$ and $U$) caused by the Zeeman effect responses to the magnetic field component perpendicular to the line of sight and is proportional to the second derivative of Stokes $I$, $d^2 I / d\nu^2$ (Crutcher et al. 1993). We waive linear polarization Zeeman effect because of its hard detectability.

Several previous attempts to detect the CCS Zeeman splitting have been
done, although no clear detection has been reported so far (Shinnaga et al. 1999b, Levin et al. 2001).
In these previous measurements, the circularly-polarized components were
measured directly with the dual circular-polarization receivers
(Shinnaga et al. 1999a, Levin et al. 2001), for
which careful polarization calibration is crucial to separate the
instrumental polarization components from the obtained signals.
Here, we developed a dual linear-polarization receiver system for which
the contribution of instrumental polarization components to the Stokes
$V$ spectrum is small, compared to the circular polarization system 
(e.g., Heiles et al. 2001, 2004). 
Such system has been used for H{$\,${\sc i}} and OH Zeeman observations at the Arecibo 
telescope (Heiles et al. 2001, Heiles et al. 2004) and detected the Zeeman splitting
of H{$\,${\sc i}} and OH successfully.

%

\subsection{System Requirements of the Receiver}

Our primary goal is to detect the Zeeman splitting of the CCS emission line at 45.3 GHz with a concrete significance ($>5\sigma$) under the magnetic fields of $\sim 100 \ \mu$G. This introduces the science requirement to measure the Stokes $V$ spectrum caused by the Zeeman splitting of 64 Hz. Our pilot observation toward TMC-1 showed the line edge profile of 
\begin{equation}
\frac{dI}{d\nu} = 0.16 \ {\rm mK \ Hz^{-1}} 
\end{equation}
over $\sim 20$ kHz that yields the Stokes $V$ of 10 mK while Stokes $I \simeq 2.5$ K (Kameno et al. 2014; 
Mizuno et al. 2014). To detect it within on-source time of 100 ks, the system noise must meet 
\begin{equation}
T_{\rm sys} < \left( \frac{10 \ {\rm mK}}{\rm SNR = 5} \right) \sqrt{2 \times 20 \ {\rm kHz} 
\times 100 \ {\rm ks}} = 126 \ {\rm K}  .
\end{equation}
Therefore, we defined the system requirements of the receiver as:
\begin{itemize}
\item[1.] Dual polarization capability to measure Stokes $V$ together with Stokes $I$.
\item[2.] Systematic error of Stokes $V$, $\Delta V$, must meet 
\begin{equation}
\frac{\Delta V}{I} < 8 \times 10^{-4} .
\end{equation}
\item[3.] The frequency fluctuation, $\Delta \nu$, should not be greater than 1/10 of expected Zeeman shift of 64 Hz, or the frequency stability of $\Delta \nu / \nu < 1.3 \times 10^{-10}$ during typical observing time of 8 hours.
\item[4.] The system noise temperature should be below 126 K at 45.3 GHz. Since the beam propagation optics of the Nobeyama 45-m telescope adds noise temperature of 30 $-$ 40 K, the receiver noise temperature must be as low as 50 K.
\end{itemize}

We examined which of linear or circular polarization reception is suitable to meet the system requirements, and concluded to adopt linear polarization system by following reasons.

In the linear polarization system, the transfer function of Stokes $V$ is described as 
\begin{equation}
V = {\rm Im} \frac{\left< XY^* \right>}{G_X G^*_Y} ,
\end{equation}
where $X$ and $Y$ stand for two orthogonal linear polarization components and $G_X, G_Y$ are the voltage-domain complex gain of the receiving system. It is remarkable that parallel-hand correlations of $\left< XX^* \right>$ and $\left< YY^* \right>$ do not responses to Stokes $V$. Since receiver noise and atmospheric radiation do not correlate between $X$ and $Y$, fluctuation of receiver gain or atmosphere will not cause significant systematic errors. The systematic error in Stokes $V$ is caused by calibration error of relative phase between $G_X$ and $G_Y$ (hereafter, $XY$ phase) and the real part of $\left< XX^* \right>$  that origins linear polarization of the target source or instrumental polarization. When we have the $XY$ phase calibration error of $\Delta \phi$ and linear polarization component of Stokes $Q$ and $U$, the systematic error in Stokes $V$, $\Delta V$, will appear as 
\begin{equation}
 \Delta V = ( (D_X + D^*_Y) I + U \cos 2\psi - Q \sin 2\psi) \sin \Delta \phi ,
\end{equation} 
where $D_X $ and $D_Y$ (D-terms) are voltage-based cross talk between $X$ and $Y$ polarization and $\psi$ is the parallactic angle. We got a perspective to achieve the required accuracy of $\Delta V / I < 8 \times 10^{-4}$ by calibrating D-terms and $XY$ phase with required accuracies of 0.01 and 0.01 rad, respectively.

For circular polarization system, the transfer function is 
\begin{equation}
V = \frac{\left< RR^* \right>}{G_R G^*_R} - \frac{\left< LL^* \right>}{G_L G^*_L} , 
\end{equation}
where $R$ and $L$ stand for right-hand and left-hand circular polarization components, respectively.
Since parallel-hand correlations directly response to Stokes $V$, calibration error in gain factor directly contributes to systematic error of Stokes $V$.
We are not confident of calibrating the differential gain, $\Delta |G_R - G_L|$, with the required accuracy of $8 \times 10^{-4}$ in single-dish observations where source emission is affected by time fluctuation of atmospheric attenuation and radiation.

To measure Stokes $V$ with a linear-polarization system, a cross-correlation spectrometer is necessary to acquire $\left< XY^* \right>$.
Recent advances in digital signal processors allowed us to build a software-based polarization spectrometer, PolariS, with a reasonable cost  (Mizuno et al. 2014). Installation of PolariS yields not only Stokes $V$ but also full-Stokes polarimetry.

In order to keep coherence between two polarization signals and frequency stability, all of the LO signals should refer to a common hydrogen-maser frequency standard. The 1st LO signal should be shared in two polarizations to avoid difference of frequency. To maintain the frequency stability throughout the signal path, we employed the VLBI backend including IF transmitter, IF downconverter, baseband converter, and digitizer. Since calibration of $XY$ phase is crucial for accuracy in Stokes $V$, a phase calibration unit composed by a wire grid should be equipped in front of the receiver feed to deliver linearly polarized in-phase waves into $X$ and $Y$.

\subsection{Target Prestellar Cores}

Since the Zeeman splitting of prestellar cores is expected to be subtle, careful target selection is also 
important.
Target cores should have strong peak intensities and narrow line widths.

The critical magnetic field strength for the dynamical stability 
is about 100 $\mu$G for cloud cores with densities of $10^4$ cm$^{-3}$. 
This causes the splitting of about 60 Hz for the CCS ($J_N=4_3-3_2$) line. 
The typical velocity width of prestellar cores is about 0.5 km s$^{-1}$
or less ($\Delta V \lesssim $ 0.5 km s$^{-1}$ $\approx 76$ kHz at 45 GHz).
In nearby star-forming regions such as Taurus, several prestellar cores
have strong CCS emission whose peak antenna intensity is greater than 1 K
($T_{\rm A, peak}^* \gtrsim $ 1 K). 
To detect the magnetic field of 100 $\mu$G within about 30 hours, the target cores
should satisfy the following two criteria: (1) the peak intensity should be
stronger than 1 K in the antenna temperature scale, and (2) the line width
should be as narrow as 0.5 km s$^{-1}$, or the line profile should have
steep envelopes so that fitting of the Stokes $V$ spectra, the first derivative
of the Stokes $I$ spectra, is easier.
We note that we apply the Smoothed Bandpass Calibration (SBC) 
method developed by Yamaki et al. (2012) for the Zeeman observations. 
The SBC allows us to reduce the total observation time by a factor of a few.

Some properties of several target cores that meet the above criteria
are listed in table \ref{tab:taurus}.
The positions of the cores in Serpens South and L1495B are shown 
in figures \ref{fig:serps} (a) and \ref{fig:serps} (b), respectively.
The CCS peak position of L1495B appears not to be accurate from the observations
of Hirota et al. (2004) because of coarse grid.
Therefore, we observed L1495B using the existing 40-GHz receiver S40 and spectrometer AC in an
on-the-fly mode, and determined the position.
The central dense part of L1495B appears round but the envelope has a
large asymmetry. 
The positions  of the cores in Serpens South is determined from the
CCS integrate intensity map presented by Nakamura et al. (2014). 
These cores are our targets of Zeeman measurements.

\subsection{Other Sciences}

Using the new receiver system, other scientific observations are also feasible.
For example, SiO masers from evolved stars or star-forming
regions often show strongly-polarized emission (Herpin et al. 2006). 
The Stokes parameters derived 
from the masers give a clue to understand the magnetic field structure in the ambient gas
around the objects. Since the SiO masers from evolved stars are often
strong, it is possible to derive full-Stokes spectroscopy from the
observed emission.
The CH$_3$OH ($\approx$ 44.1 GHz) maser lines also show the Zeeman effect, 
which is recently detected toward OMC-2 (Sarma \& Momjian 2011).
Our receiver has a capability to detect the Zeeman effect of this CH$_3$OH line.
The CH$_3$OH maser lines arise from shocks in the protostellar outflows, 
and thus we can gain information of the magnetic fields associated with 
the gas shocked by the protostellar outflows. 
We can also obtain information of the magnetic field structures around the AGN jets. 
CO ($J=3-2$) line from high redshift galaxies can fall into this band,
and if detected, we can derive the redshifts of high-z galaxies accurately.

\section{Development}
\label{sec:develop}

To conduct the Zeeman observations, we developed a new 40 GHz receiver 
of dual polarization. 
In the following, we  describe the details of the receiver, focusing on
optics, horn, and OMT (Ortho-Mode Transducer).

\subsection{Optics and Horn}

The NRO 45-m telescope adopts a beam-guide antenna design.
The optics of the telescope is briefly described in 
figure 1 and table 1 of Nakajima et al. (2008).

\begin{figure}
	   \begin{center}
  \includegraphics[height=6cm]{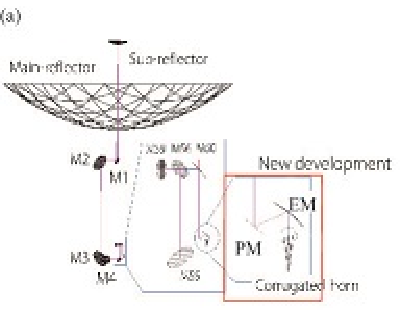} 
  \includegraphics[height=6cm]{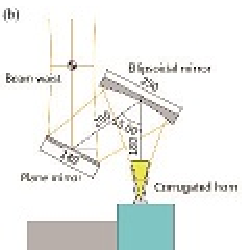} 
 \end{center}
\caption{(a) Optics and the beam propagation design of the NRO 45-m telescope. 
PM and EM denote the plane and ellipsoidal mirrors, respectively.
The developed 40-GHz band receiver system enclosed with the red box is installed in the receiver
 cabin.  A signal from astronomical object focused by a Cassegrain
  system is reflected by a plane mirror, M1, and is led to a receiver
  system by a pair of mirrors (M2 and M3).
(b) Blow-up of the red box in Fig.\ref{fig:mirror}a.}\label{fig:mirror}
\end{figure}

To install the new receiver system in the NRO 45-m telescope, 
we developed a plane mirror, an ellipsoidal mirror, and a corrugated horn. 
We used the Gaussian optics to design the optics for the new receiver.
Here, we focus on optics of the new receiver.

Using the GRASP9 software, we simulated the beam pattern with the physical optics.
Assuming the edge taper levels of the sub-reflector and mirrors to be 
$-$12 dB and $-$34 dB, respectively, we designed the new optics based 
on the Gaussian beam propagation.
We did not take into account the effects of blocking by the 
sub-reflector stays, surface errors of mirrors, and ohmic loss of the optical elements.
In the GRASP9 calculations, we assigned the optical components such as the horn and mirrors in the virtual 
three-dimensional space, and divided their surfaces into small elements. Then, we calculated
the electric fields at the elements on the horn surface by numerically solving Maxwell's equations, 
and repeatedly calculated the electric fields on the surface of the next optical component 
so that we obtained the electric field of the designed optical system.

\begin{figure}
	   \begin{center}
  \includegraphics[width=16cm]{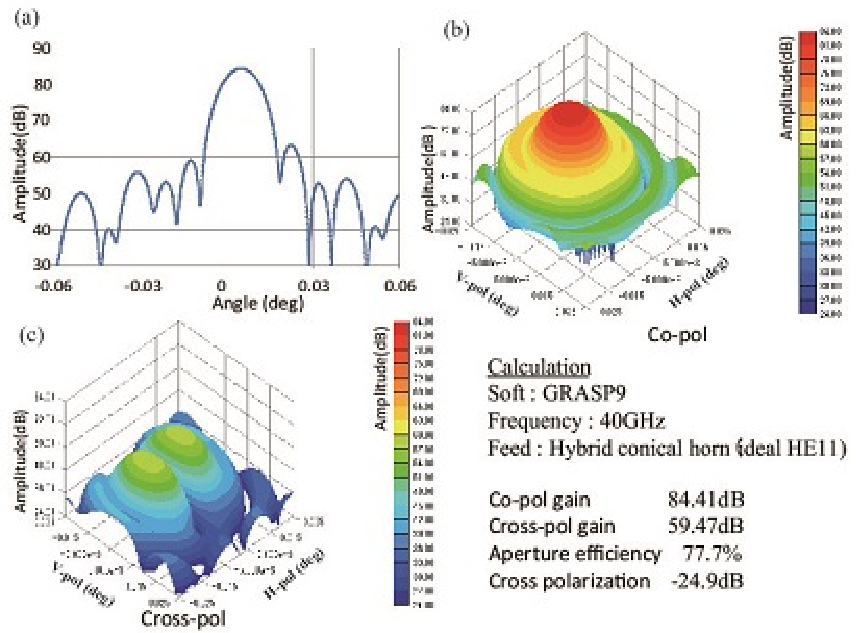} 
 \end{center}
\caption{(a) Simulated beam pattern at 40 GHz on the V-polarization plane, measured when the elevation is equal to 90$^\circ$. The horizontal axis corresponds
to the angle measured from the vertical direction of the V-polarization plane. Note that the 45-m optical system has an offset of 1.06585$^\circ$ from the Cassegrain focus axis
and thus the beam pattern has a peak slightly offset from angle=0$^\circ$.
(b) Bird's eye view of the simulated beam pattern and (c)
Bird's eye view of the cross-polarization  pattern. H-pol and V-pol denote the horizontal and vertical polarization directions, respectively.
We used the GRASP9 physical optics software to design the optics assuming that the elevation of the telescope is equal to 90$^\circ$.}
\label{fig:beam}
\end{figure}

In figure \ref{fig:mirror}, we show the design of the  beam propagation 
of our developed system at 40 GHz. We also assumed the feed to be a hybrid conical
horn (ideal HE11) in the simulations, because we 
designed the corrugated horn to have the beam pattern comparable to that of a hybrid conical horn.
We have reduced the spillover loss by increasing the size of the mirrors as large
as possible. In addition, we tried to minimize the beam bend angle at the ellipsoidal
mirror in order to reduce the effect of the cross polarization coupling. Figure
\ref{fig:beam} shows the results of the simulations.  The parameters of the Gaussian optics 
used for the simulation are summarized in table \ref{tab:optics}.
The cross-polarization characteristics was computed to be $-24.9$ dB.
The co-polarization and cross-polarization gains were estimated to be
84.4 dB and 59.5 dB, respectively.
An antenna directivity was estimated to be  84.4 dB, corresponding to
an aperture efficiency of 0.78 for the NRO 45-m telescope.
The first side-lobe level was calculated to be $-20$ dB.

\begin{table}
  \caption{Parameters of the Gaussian optics used for the simulation. The dimensions are given in mm, otherwise indicated.}
\label{tab:optics}
  \begin{center}
    \begin{tabular}{llll}
     \hline
      Separation between mirrors & & & \\
      Sub-ref to M1 (plane)& 20211.9 & Beam waist $\omega_{0, \rm SR}$ & 30.63 \\
      M1 to M2 (ellipsoidal) & 6500 & Beam size at M2 & 486.13  \\ 
      M2 to M3 (ellipsoidal) & 17000 & Curvature at M2 & 6255.2 \\ 
      M3 to M4 (plane) & 2300 & Curvature at M2 (image) & $-$965259.72 \\ 
      M4 to M55 (plane) & 3000 & Beam waist $\omega_{0, \rm M2}$ & 483.59  \\ 
      M55 toM56 (ellipsoidal) & 2500 & Beam size at M3 & 501.68  \\
      M56 to WG (plane) & 1200 & Curvature at M3 & 38172.06  \\
      WG to M60 (ellipsoidal) & 1200 & Curvature at M3(image) & 6401.46 \\
      M60 to EM (ellipsoidal) & 2300 & & \\
      EM to currugated horn & 180 & Beam waist $\omega_{0, \rm M3}$ & 30.39 \\
      & & Beam size at M56 & 115.71  \\
      Focus distance of ellipsoidal mirrors & & Curvature at M56 & 1527.35 \\
      M2 & 6296 & Curvature at M56(image) & $-$23607.81 \\
      M3 & 6296 & Beam waist $\omega_{0, \rm M56}$ & 112.57 \\
      M56 & 1633 & Beam size at M60 & 136.74  \\
      M60 & 1633 & Curvature at M60 & 11366.33 \\
      EM & 194.87 & Curvature at M60(image) & 1906.98 \\
       &  & Beam waist $\omega_{0, \rm M60}$ & 32.33 \\
      Edge taper of sub-ref & 12 dB & Beam size at EM & 49.03 \\
      Diameter of sub-ref & 3750 & Curvature at EM  & 883.7\\
      Beam size at sub-ref  & 1595.21 & Curvature at EM (image) & 250 \\
      Curvature at sub-ref & 20489.1 & Beam waist $\omega_{0, \rm EM}$ & 11.81\\
      Diameter of horn aperture & 50.613 & Beam size at $\omega_{\rm horn aperture}$ & 16.28\\
      Slant length of horn & 116.99 &  Curvature at horn aperture& 116.99 \\
      \hline
    \end{tabular}
  \end{center}
  The subscript SR means sub-reflector.
\end{table}

\begin{figure}
	   \begin{center}
  \includegraphics[width=10cm]{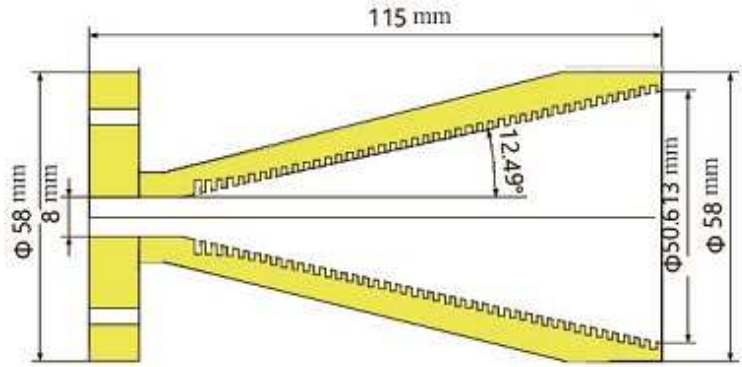} 
 \end{center}
\caption{Cutaway view of the corrugated horn}
\label{fig:horn}
\end{figure}

\begin{figure}
	   \begin{center}
  \includegraphics[width=7cm]{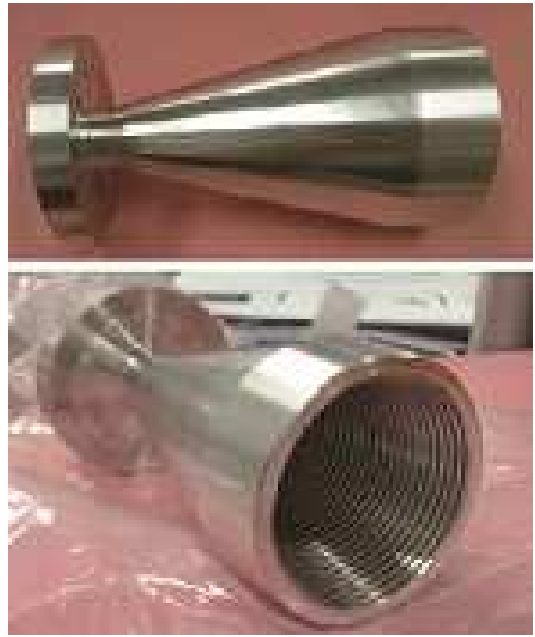} 
 \end{center}
\caption{Photographs of the developed corrugated horn}
\label{fig:hornimage}
\end{figure}

\begin{figure}
	   \begin{center}
    \includegraphics[width=8cm]{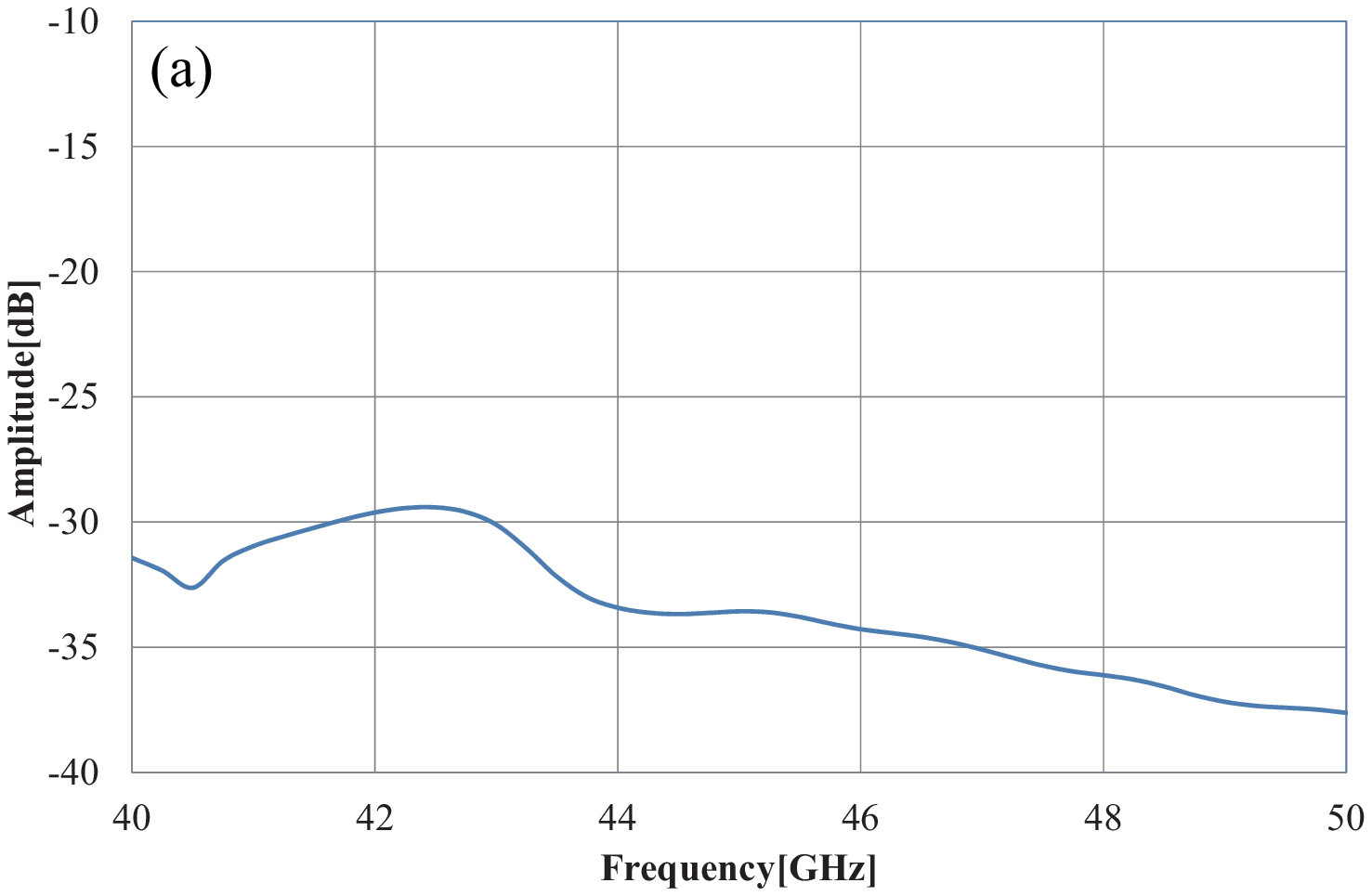}
    \includegraphics[width=8cm]{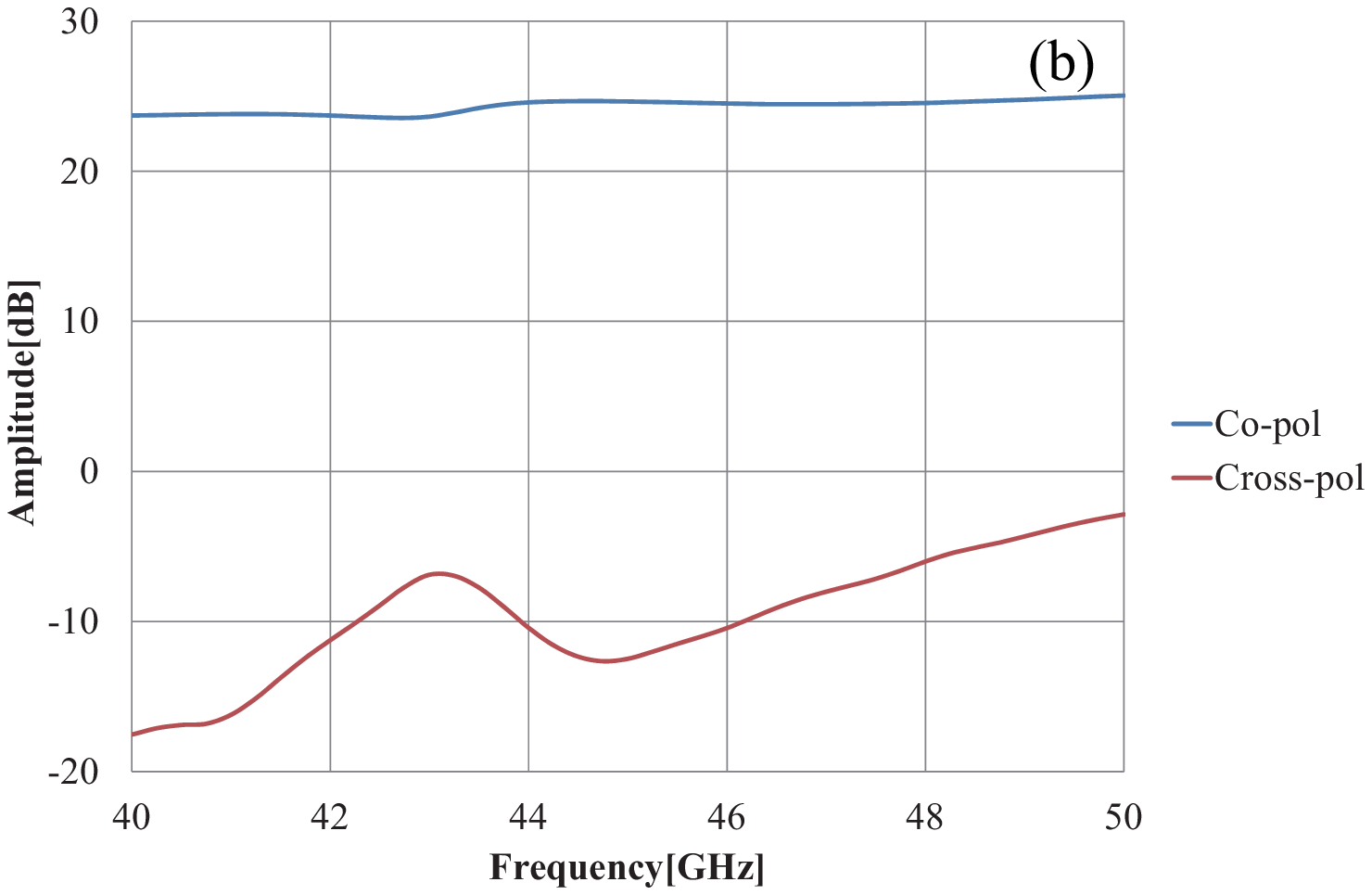}
    \includegraphics[width=8cm]{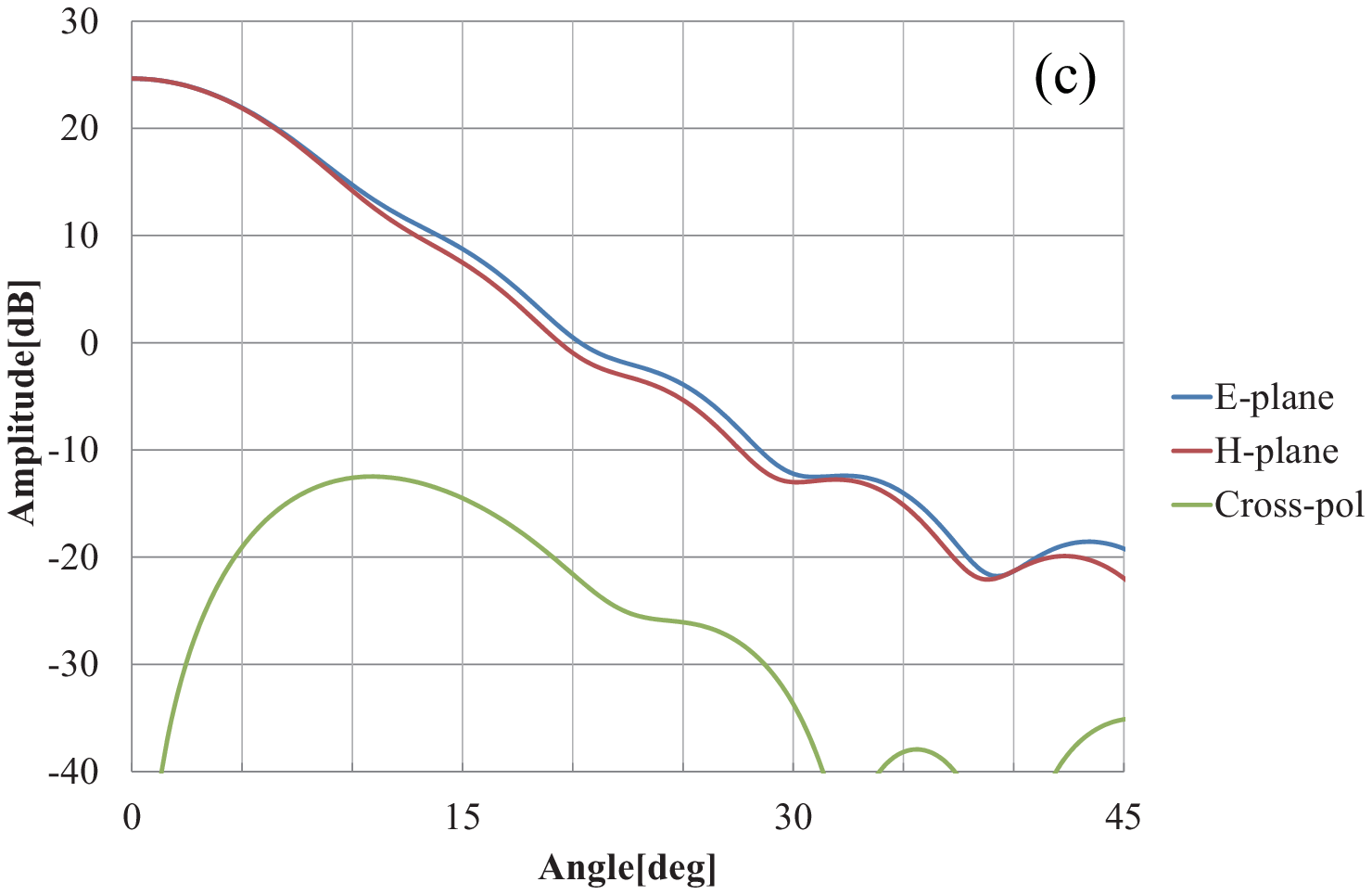}
 \end{center}
\caption{(a) Simulated return loss, (b) peak  directivity as a function
 of frequency, and (c) beam
 profile of the corrugated horn at 40 GHz, as a function of angle. 
In panel (b), the blue and red curves denote
the main-polarization and cross-polarization components, respectively.
In panel (c), the blue, red, and green curves denote the E-plane,
 H-plane, and cross-polarization components, respectively.\label{fig:horn2}}
\end{figure}

Corrugated horns are widely used as a feed horn for microwave antenna
systems. This is because of their superb radiation performance. In particular,
their high co-polar pattern  symmetry and low cross polarization are best
in our polarization observations.
We repeatedly calculated a suitable corrugation pattern from
the basic design using CHAMP and finally determined the design presented in
figure \ref{fig:horn}. 
The physical dimensions of the corrugated horn are presented in table
\ref{tab:horn}. 
In figure \ref{fig:hornimage} we also present the photographs of the developed 
corrugated horn.
The width of the grooves is 1.09 mm and the total number of grooves is 43.
According to  our calculation, a return loss was estimated to be at most 25 dB.
A maximum cross polarization level lower than $-$35 dB from the peak
 were achieved
(figure  \ref{fig:horn2}).
The simulated characteristics of the horn are presented in Fig. \ref{fig:horn2}
which shows the similarity of the bean patterns between E-plane and
H-plane down to $-$20 dB. These beam patterns are in good agreement with
the theoretical beam pattern obtained by the radiation of the HE11 mode.
We built the horn by the directly-dig method (Kimura et al. 2008).

\begin{table}
  \caption{
Physical dimensions of the corrugated horn}\label{tab:horn}
  \begin{center}
    \begin{tabular}{ll}
     \hline
Material & Aluminum \\
Total length & 115.00 mm \\
Length of the corrugation section & 100.00 mm \\
Outer diameter of the aperture & 58.00 mm \\
Inner diameter of the aperture & 50.613 mm \\
Semi-flare angle & 12.49$^\circ$ \\
Diameter of the circular waveguide & 8.00 mm \\
Width of the grooves & 1.09 mm \\
Depth of the grooves at the mode-launching section & 2.86 mm \\
Depth of the grooves at the thread section & 1.66 mm \\ \hline
\end{tabular}
\end{center}
\end{table}

\subsection{Receiver}
\label{sec:receiver}

\begin{figure}
	   \begin{center}
  \includegraphics[width=8cm]{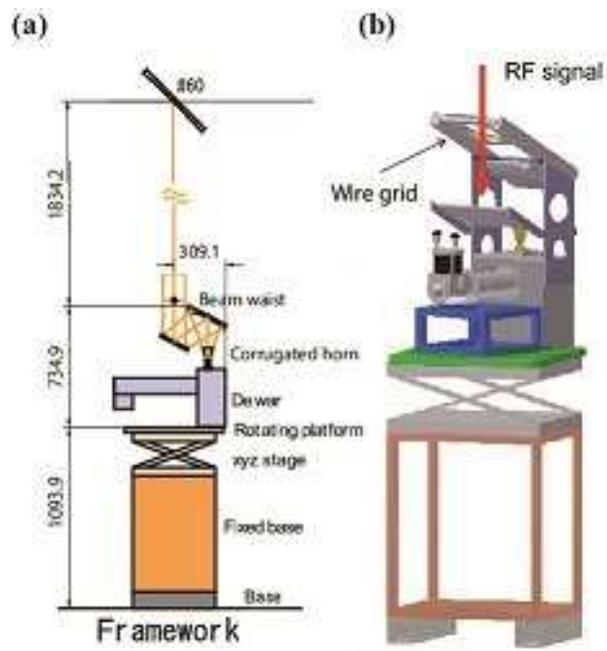} 
 \end{center}
\caption{3D overall view of the receiver system}
\label{fig:receiver1}
\end{figure}

\begin{figure}
	   \begin{center}
  \includegraphics[width=10cm]{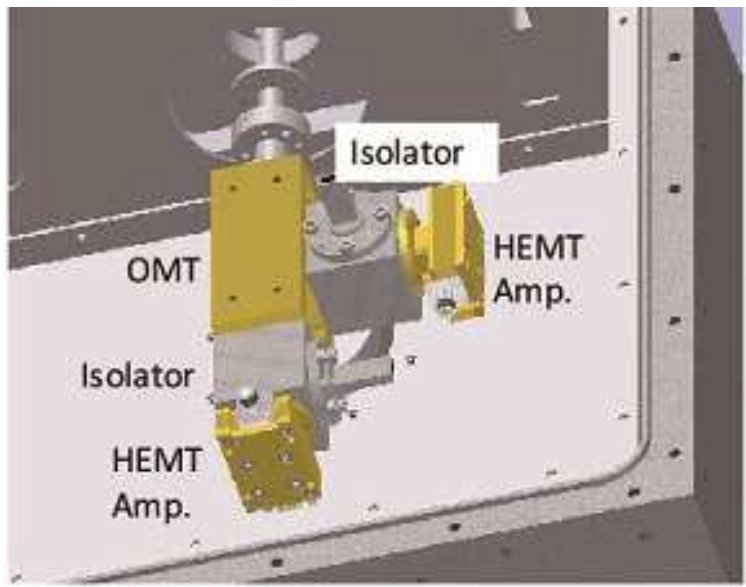} 
 \end{center}
\caption{A schematic view in the dewar.}
\label{fig:dewar}
\end{figure}

\begin{figure}
	   \begin{center}
\includegraphics[width=12cm]{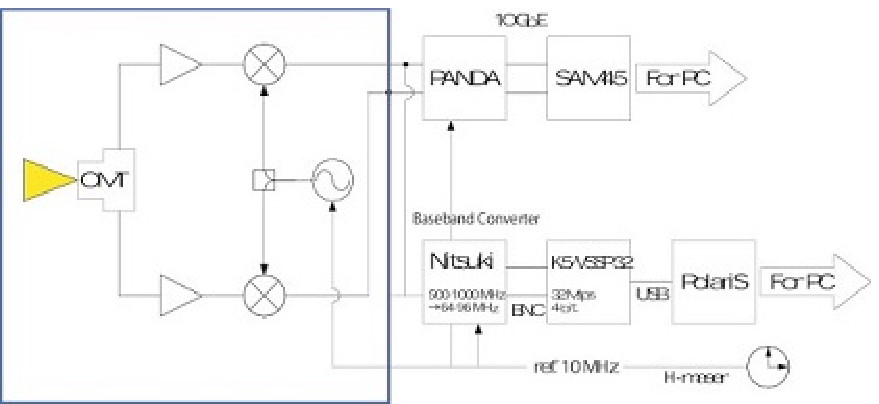} 
\includegraphics[width=12cm]{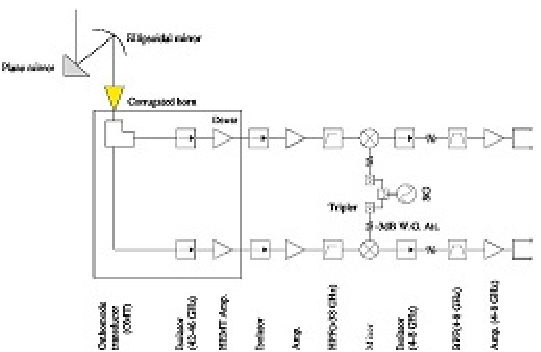}   
 \end{center}
 \vspace{2cm}
\caption{({\it upper}) Block diagram of the new 40 GHz receiver system.
({\it bottom}) Enlargement of the blue box in the upper panel.}
\label{fig:blockdiagram}
\end{figure}

We have designed the cooled receiver whose schematic picture is
shown in figure \ref{fig:receiver1}.
The receiver consists of base, rotating platform, dewar, and
calibration unit. The receiver is installed on the rotating platform
to perform polarization calibration. It also has the calibration unit
on the upper side of the receiver system.
The design of the dewar is presented in figure \ref{fig:dewar}. 
The dewar is cooled down to about 20 K.
The HEMT amplifiers are connected to OMT with isolators.
The specifications of the receiver system is summarized in table \ref{tab:receiver}. 
A block diagram of the receiver system is shown in figure \ref{fig:blockdiagram}.
The incoming signal is brought to a corrugated horn using plane and
ellipsoidal mirrors. The signal is split into two linearly-polarized
components via OMT.  Then, each component is amplified  and
down converted with a local signal with a frequency tripper using the upper sideband mixing scheme.
The operation frequency range is from 42 to 46 GHz, which is limited by
the frequency range of the local oscillator (LO) chain.
IF frequency range is  4.0-8.0 GHz with a bandwidth of 4.0 GHz.
The signal is amplified by a HEMT amplifier (Nitsuki 9837QC) whose
noise temperature is around 29 K at 45 GHz.
The HEMT amplifiers used are cryogenically-cooled
InP monolithic microwave integrated circuits (MMIC's) made by AMMSys Inc, designed for the VSOP2 project, 
and their details are described in Nakano et al. (2012).

We adopt warm optics because of the following two reasons: (1)
the limited cost of the receiver development and (2) 
the large contribution from the NRO 45-m beam transmission system to the total system noise 
temperature ($\sim 30$ K).

\begin{table}
  \caption{Specifications of the 40-GHz new receiver system}\label{tab:receiver}
  \begin{center}
    \begin{tabular}{ll}
     \hline
      Type of receiver & HEMT \\
      Optics system & warm \\
      RF frequency & 42$-$46 GHz \\
      local oscillator & signal generator with frequency tripler \\
      IF frequency & 4$-$8 GHz \\
      Polarization & dual linear polarization \\
      Receiver noise temperature & $\sim 50$ K \\
      \hline
    \end{tabular}
  \end{center}
\end{table}

The receiver system is connected to two spectrometers:
SAM45 and PolariS. SAM45 is a highly flexible FX-type 
digital spectrometer (Kamazaki et al. 2012) having 16 sets of
a 4096 channel, whose frequency resolution can reach 3.81 kHz
($\approx 0.025$ km  s$^{-1}$ at 45 GHz). It is used as a back end 
for other receivers such as H20, H40, TZ and FOREST of the NRO 45-m
telescope (see Nakajima et al 2013a, 2013b for TZ and FOREST).
Eight sets of SAM45 are connected to Z45: four of them are for the horizontal components, 
while the rest are for the vertical components.

PolariS is the software-based polarization spectrometer whose frequency 
resolution is 60 Hz.
The PolariS consists of K5/VSSP32 digitizer by Nitsuki (Kondo et al. 2006) 
and a Linux-based PC with GPU.
It is designed to process full-Stokes spectroscopy of 2 $\times $ 131072
channels for bandwidth of 4 or 8 MHz and newly developed for our project.
See Mizuno et al. (2014) for detail.
SAM45 has a capability to process the full-Stokes spectroscopy, but
has not been fully tested yet.  Therefore, we use PolariS for our Zeeman
observations.

\begin{figure}
	   \begin{center}
\includegraphics[width=8cm]{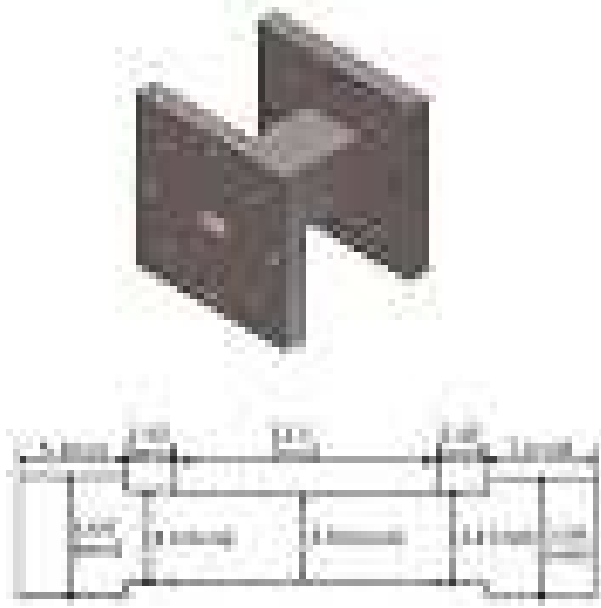}   
 \end{center}
\caption{({\it upper}) A schematic view of the high pass filter. ({\it lower}) Detailed structure of the high pass filter.}
\label{fig:highpassfilter1}
\end{figure}

\begin{figure}
	   \begin{center}
\includegraphics[width=8cm]{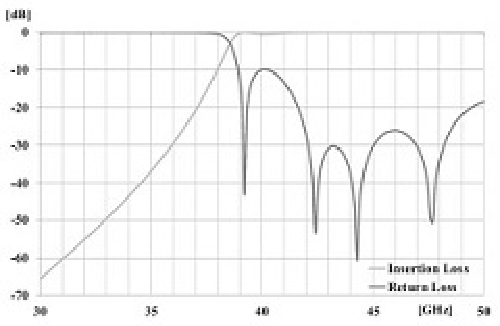}   
 \end{center}
\caption{Simulated return loss and insertion loss of the high pass filter.}
\label{fig:highpassfilter2}
\end{figure}

We also designed a high pass filter that cancels out the contribution of the lower sideband.
The cut off frequency, and the insertion loss of the LSB port
are set to 38 GHz and 30 dB,  respectively. The return loss of the USB
is set to be larger than 20 dB. The details are shown in figure \ref{fig:highpassfilter1}.
In figure \ref{fig:highpassfilter2} we present the results of the simulation. 
We also measured the performance of the developed high pass filter in the lab and 
confirmed that the return loss and insertion loss are consistent with the simulation results.

One of the unique parts of the receiver is the calibration unit, wire
grid, which is attached on the top of the optics holder (see e.g., figure \ref{fig:receiver1}). 
The wire grid is slanted off the horizontal 
with an angle of 27.5$^\circ$.
The inner diameter of the wire grid is 200 mm, so that the wire grid fully 
covers the beam near its waist.
We mounted the wire grid unit on the top of optics holder  for the calibration of
polarization observations. 
The wire grid can be mounted and demounted 
by sliding up and down. The wire grid makes the incoming flux highly-polarized 
with a certain polarization angle with respect to the receiver, so that 
we can estimate the delay and phase between two linear polarizations.

\begin{figure}
	   \begin{center}
  \includegraphics[width=15cm]{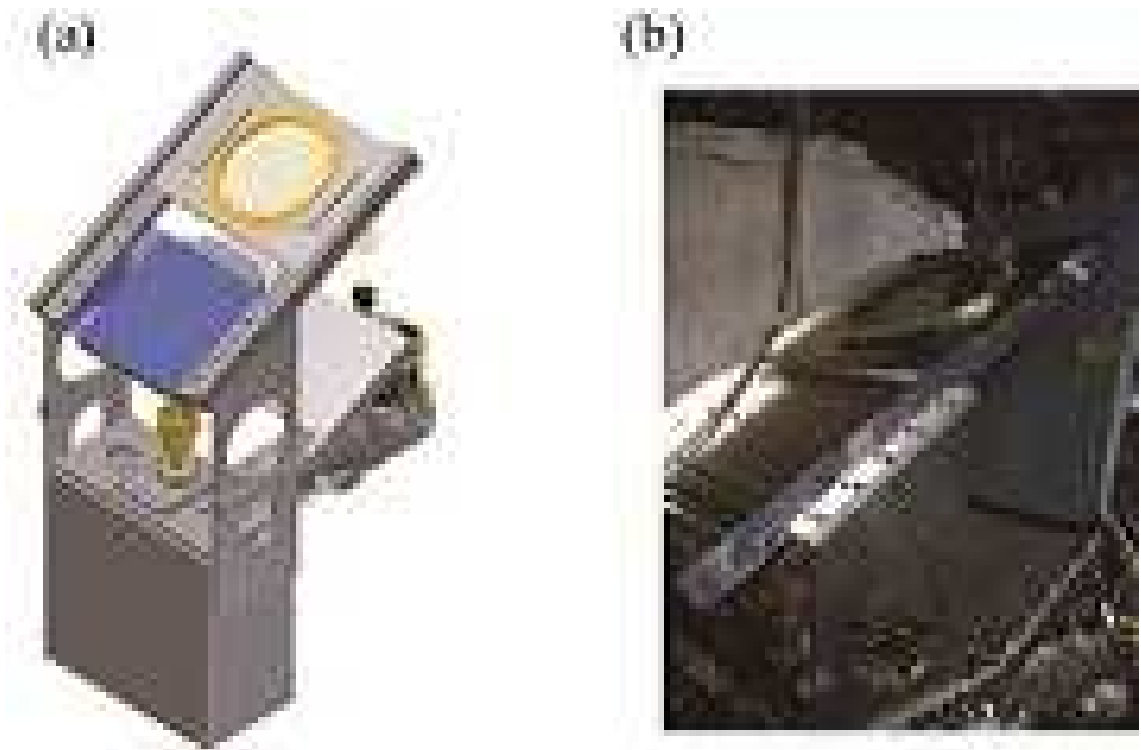} 
 \end{center}
\caption{(a) A overall view of the receiver system mounting the wire grid unit which
slides down to the blue area during the actual observations. 
(b) Photograph of the wire grid unit}
\label{fig:wiregrid}
\end{figure}

The configuration diagram of the wire grid is shown in figure
\ref{fig:wiregrid} (a).
The photograph of the calibration unit in the receiver cabin is shown in
figure \ref{fig:wiregrid} (b).
The wire grid unit is demounted for the standard observations
with the SAM45 spectrometer.
Using the wire grid, we can make the linearly polarized signals with 
polarization degree of more than 50 \%.
The details of the wire grid and calibration process using it are shown in
a forthcoming paper (Mizuno et al. in preparation).

\subsection{Ortho-Mode Transducer}

\begin{figure}
	   \begin{center}
  \includegraphics[width=12cm]{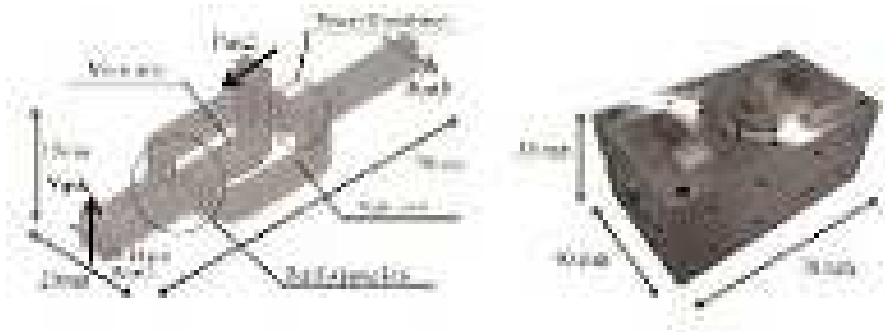} 
 \end{center}
\caption{(a) Wire frame model of the waveguide structure. The vertical
 polarization
(V pol.) goes through a further stepped transition. The horizontal
 polarization (H pol.) is split into the two side arms, and recombined by
 a power combiner. (b) A photo of the assembled OMT, which was
 manufactured from the tellurium-copper and gold plated.}
\label{fig:omt}
\end{figure}

\begin{figure}
	   \begin{center}
    \includegraphics[width=10cm]{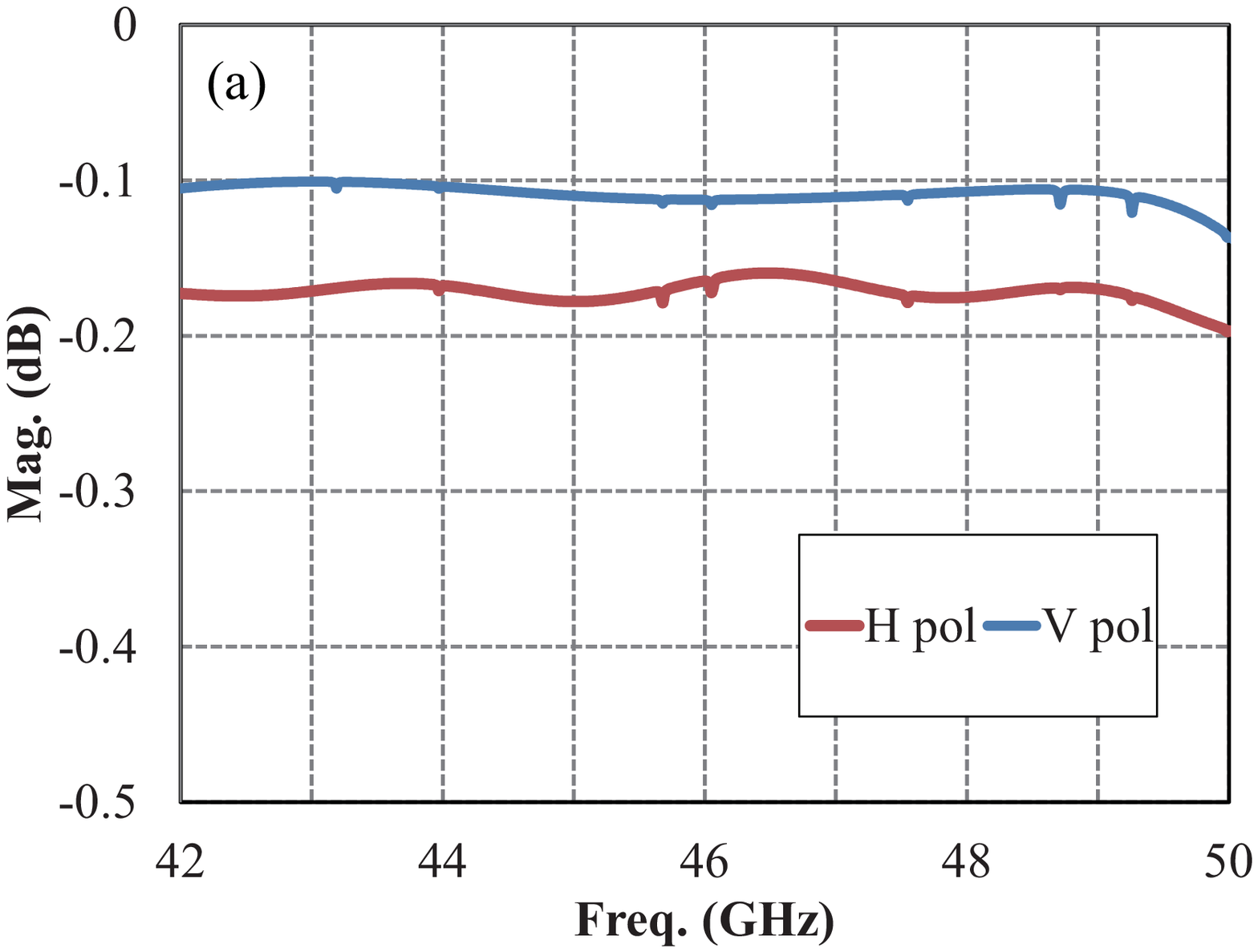} 
    \includegraphics[width=10cm]{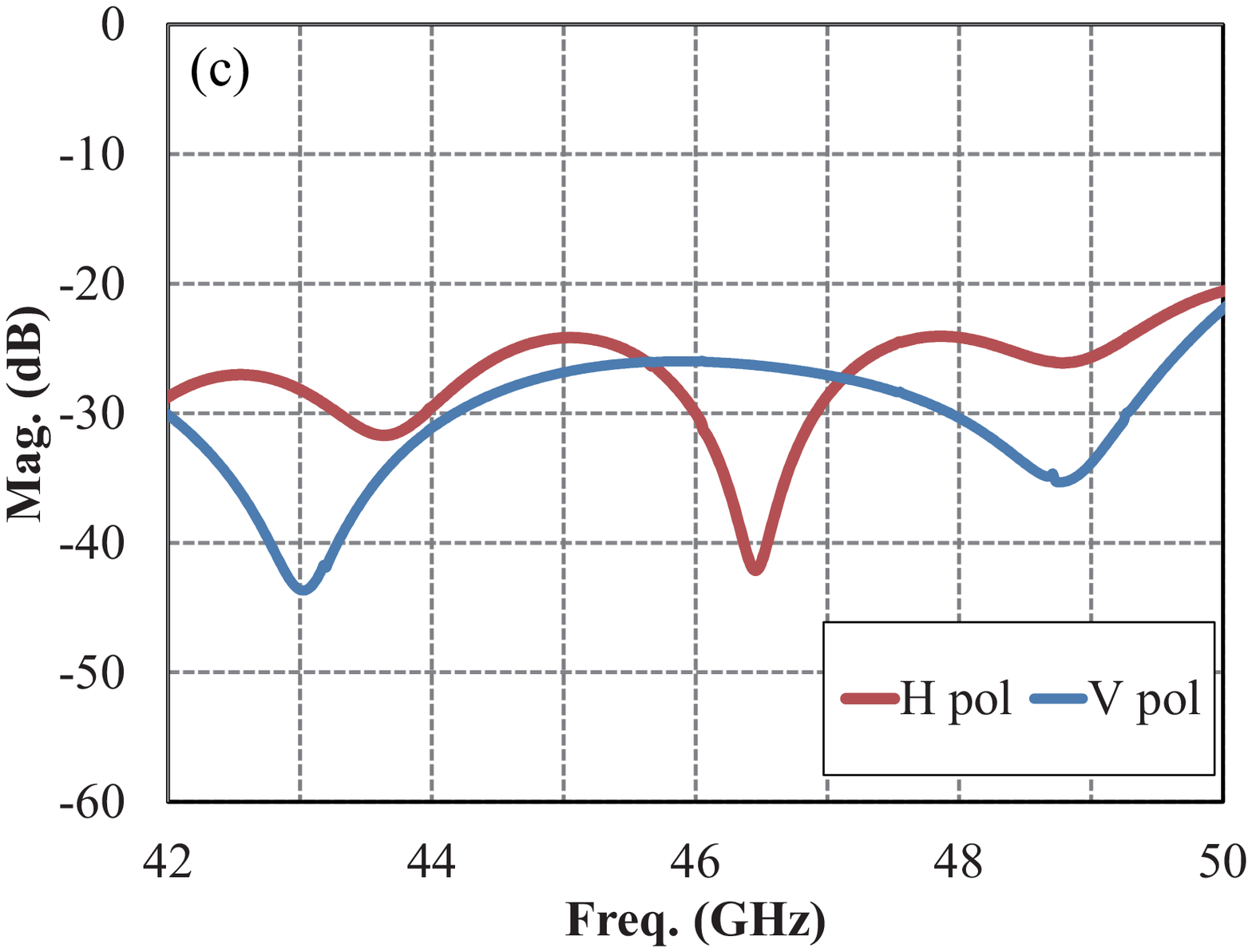} 
    \includegraphics[width=10cm]{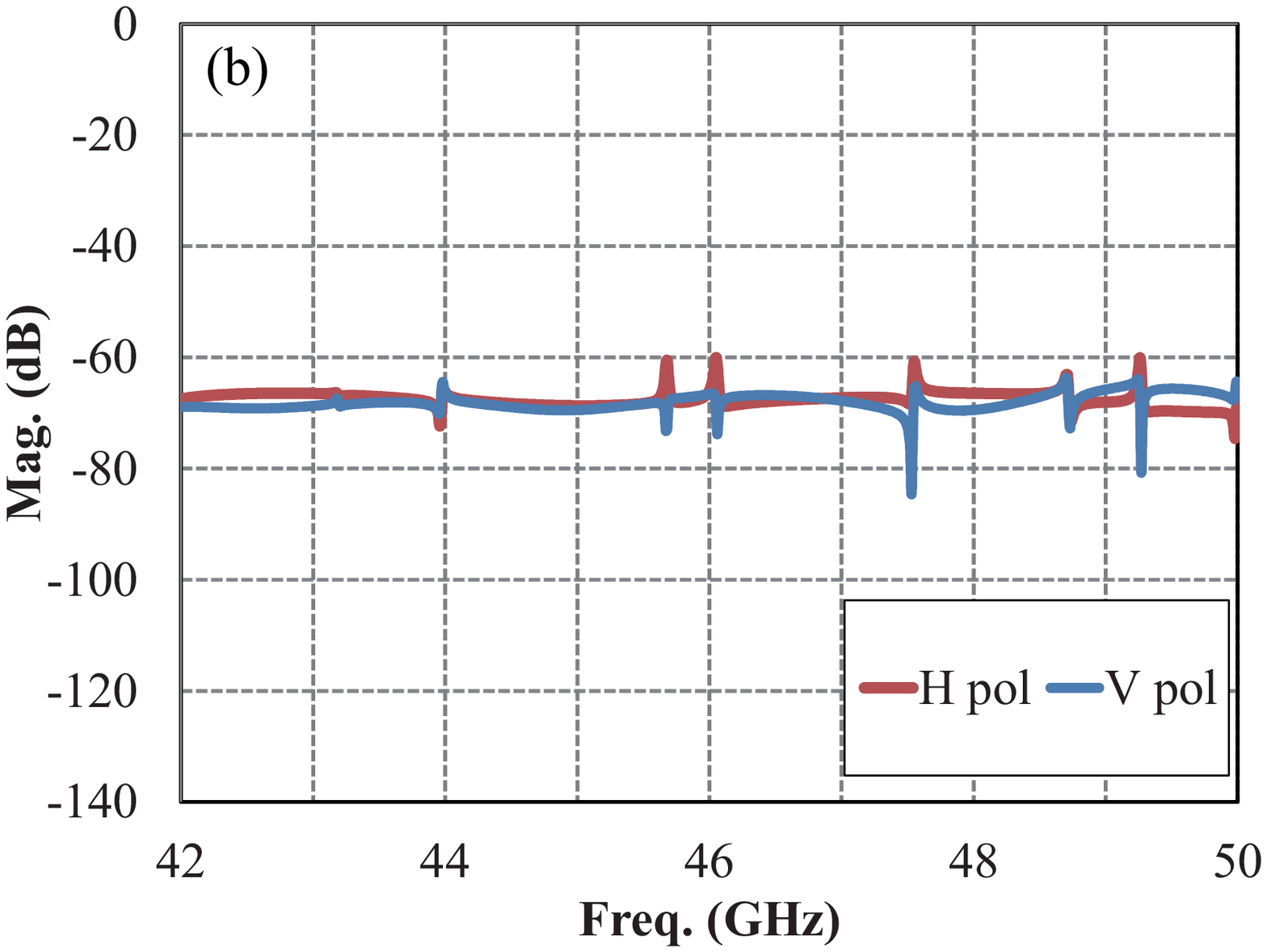} 
 \end{center}
 \caption{Simulated performance of the OMT. (a) The insertion loss, (b) return loss, and (c) cross
 polarization coupling of the OMT.}
\label{fig:omt2}
\end{figure}

To perform polarization observations, it is important to separate an
incoming signal into two orthogonal linearly-polarized components.
To do so, we developed a new double-ridged waveguide OMT for the new receiver.
The design is basically similar to that of Asayama \& Kamikura (2009).
The OMT consists of a square double-ridged wave guide transmission followed
by B$\phi$ifot junction (Moorey et al. 2006) of two side arms with the central guide.
We designed and optimized it using a 3D electromagnetic simulator (ANSYS Corporation,
High Frequency Structure Simulator  (HFSS) version 13.0).
The designed OMT is shown in figure \ref{fig:omt}. 
The incoming signal is brought from Port 1, the left edge of the OMT, and
horizontal and vertical components come out of Port 3 and 2, respectively. 
For the Q band (40 GHz) design, a 5.69 $\times$ 5.69 mm square 
waveguide is adopted for input and 5.69 $\times$ 2.84 (WR-22) 
rectangular waveguides for the outputs. 
Frequency dependence of insertion loss, return loss, and
cross-polarization coupling are shown in figure \ref{fig:omt2}.
We measured the performance of the developed OMT by using a network analyzer, and
confirmed that the OMT has a return loss of more than 20 dB, 
a cross polarization coupling of lower than $-$30 dB and an insertion loss of 
less than 0.3 dB across the observational frequency band.

\subsection{Noise Temperature}

\begin{figure}
	   \begin{center}
  \includegraphics[width=13cm]{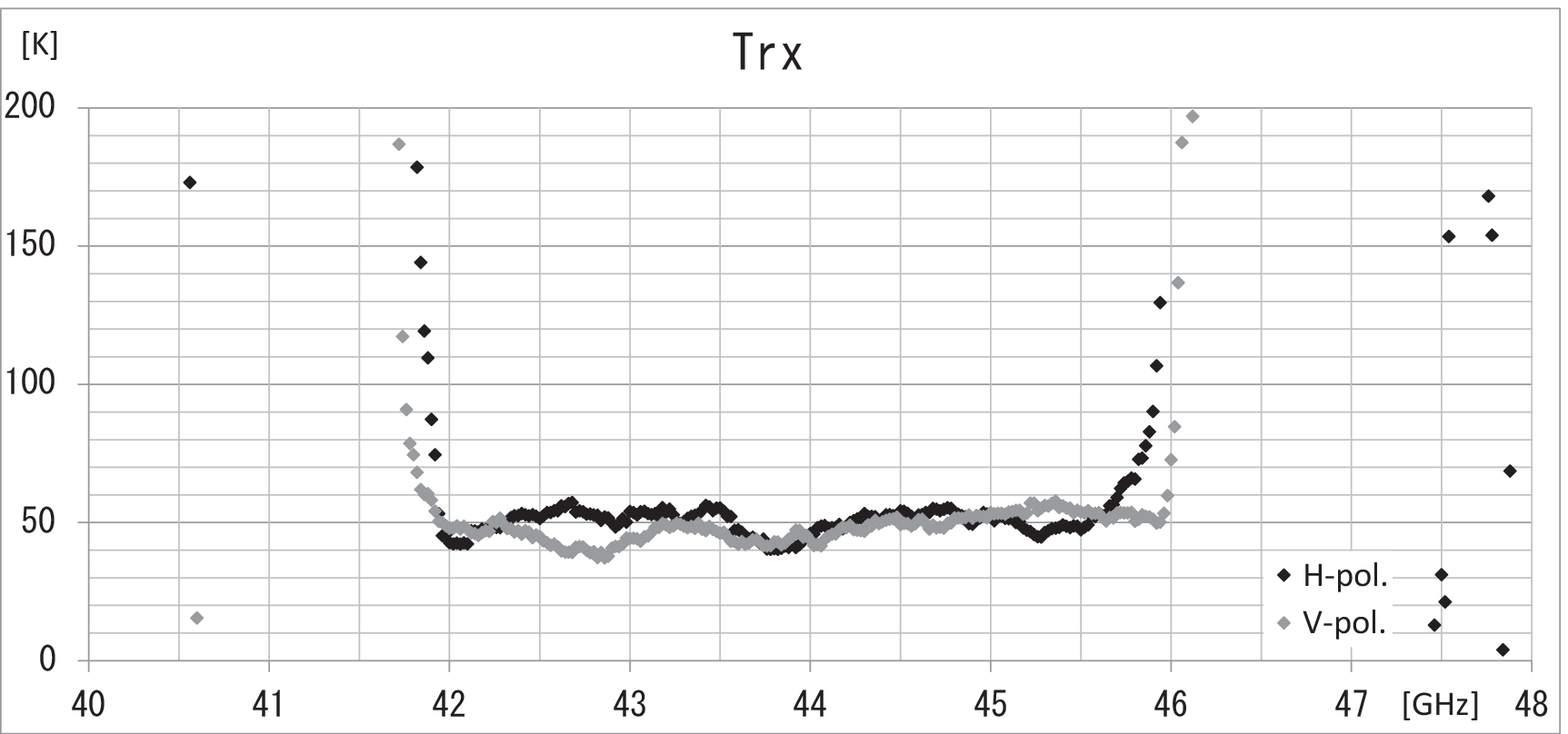} 
 \end{center}
\caption{Measured noise temperatures for the H and V polarizations.}
\label{fig:noise}
\end{figure}

To evaluate the performance of the receiver system, we measure the noise
temperature
by a standard Y-factor method using hot (300 K) and cold (77 K) loads in
the laboratory. 
Figure \ref{fig:noise} shows the measured noise temperatures for H and V polarizations
as a function of the frequency.  In the range of 42 to 46 GHz, the noise temperatures
for both H and V polarizations are equal to about 50 K, being satisfied 
with the requirement of the development.

\subsection{Installation}

The receiver system was installed in the NRO 45-m telescope in March 2013
to measure its performance, and we demounted it  in June 2013. After
adjusting some parts in the laboratory, we reinstalled it in January
2014. 
The first astrophysical observations were conducted on April 2, 2013 toward the SiO maser, 
T-Cep, whose coordinates are (R.A.[J2000.0], decl.[J2000.0])=(21:9:31.78, +68:29:27.2).


\section{Performance of the New Receiver}
\label{sec:results}

\subsection{System Noise Temperature}

The noise temperatures of the system, 
including the atmosphere, were approximately 100 K at around 43 GHz 
for H and  V polarization components at an elevation of 60$^\circ$. 
The beam propagation system of the NRO 45-m telescope adds at least a noise temperature of about 30 K
to the receiver system. Therefore, the measured system noise temperature 
is consistent with the measurement at the laboratory ($T_{\rm RX} \sim 50$ K).
This system noise temperature is smaller than those of 
the current available single-polarization 
receivers such as S40 ($\sim 150-300$ K) and H40 ($\sim 180$ K).

\subsection{Beam size, Main Beam Efficiency}

\begin{figure}
	   \begin{center}
 \includegraphics[width=8cm]{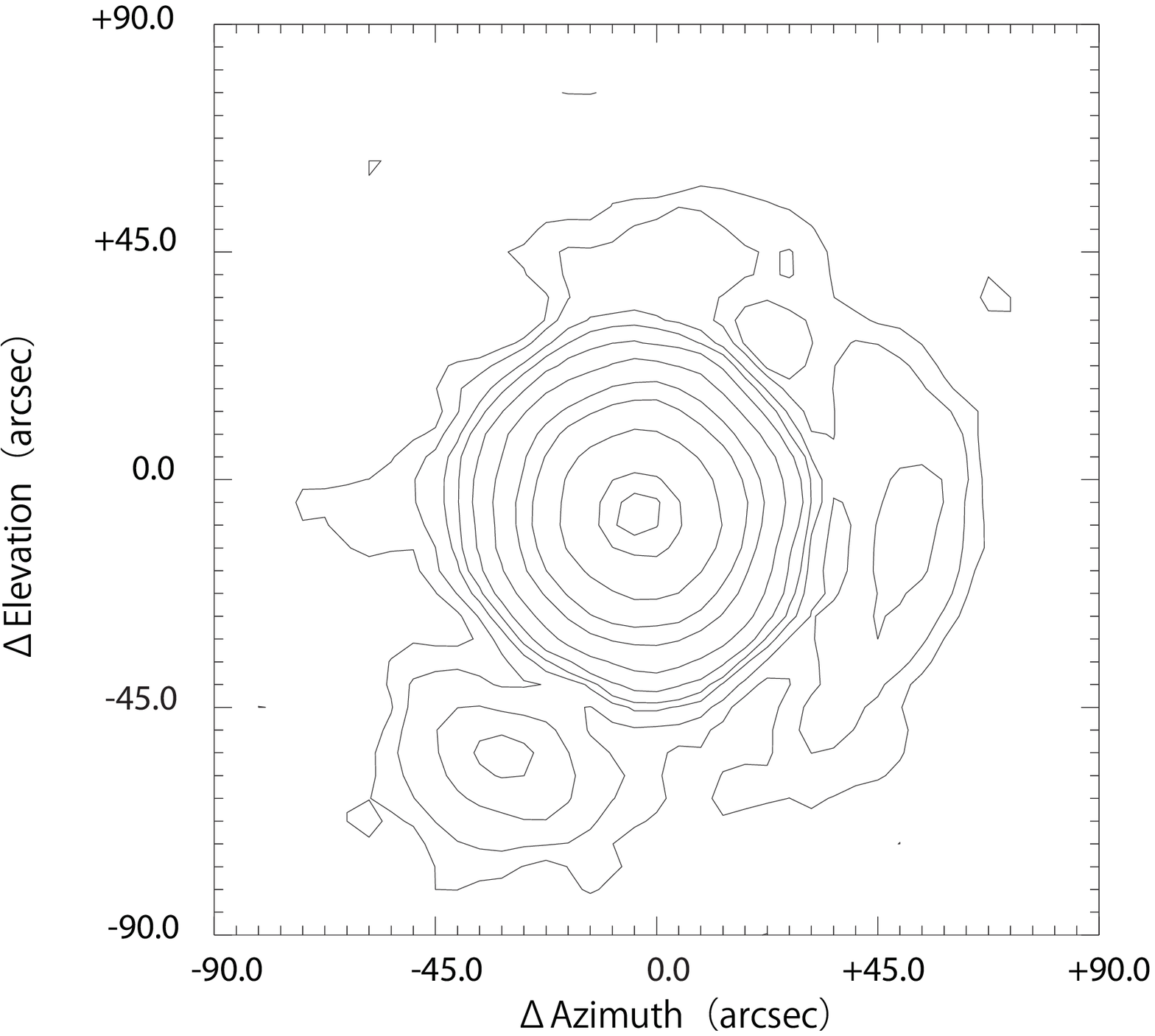}
 \includegraphics[width=8cm]{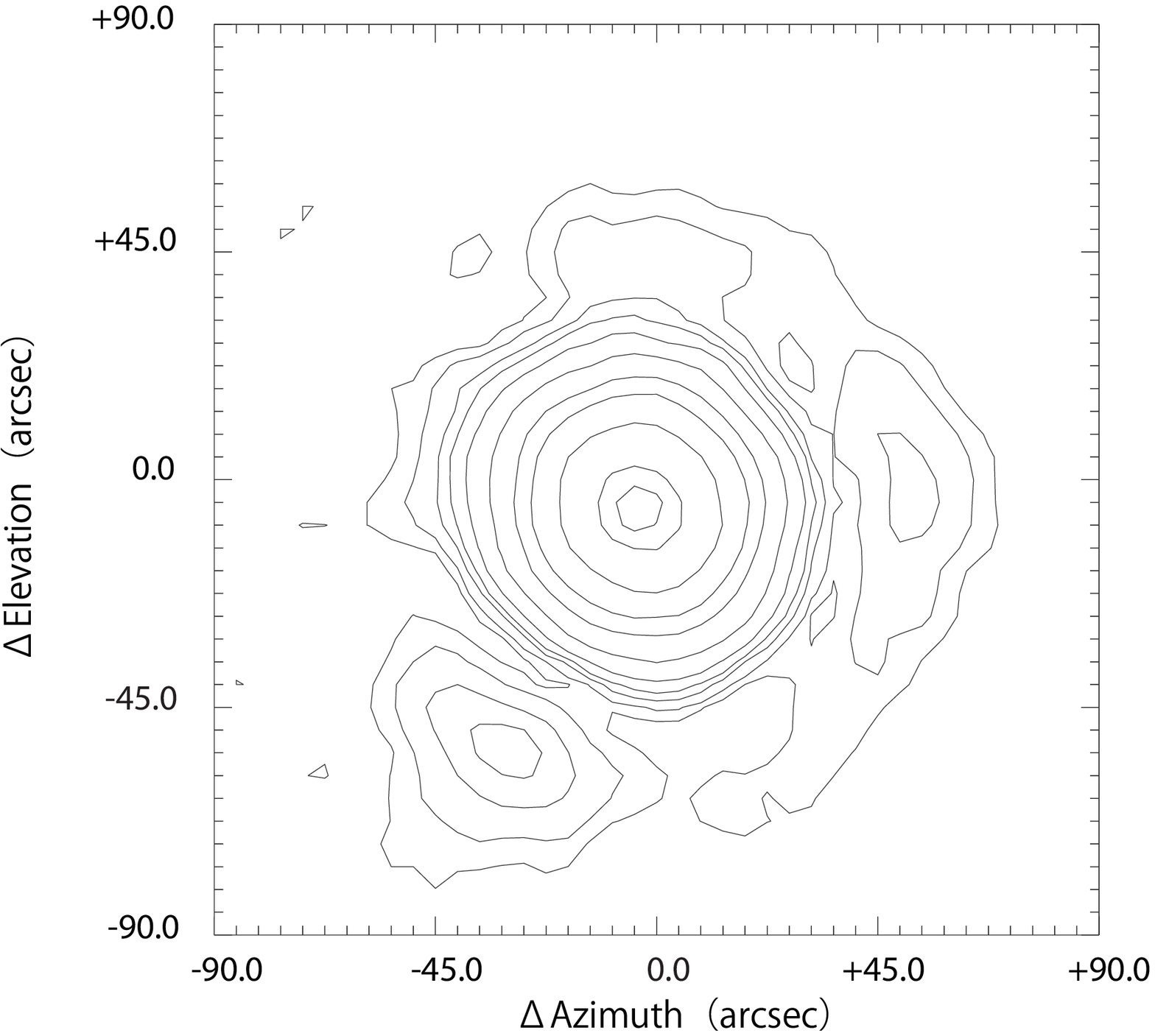} 
 \end{center}
\caption{Maps of the telescope beam constructed from SiO maser
 observations of R-Leo. Panel (a) and (b) are the integrated intensity
 of H and V-components, respectively.
The contour are drawn at 0.5, 1, 2, 3, 5, 7, 10, 20, 30, 50, 70, 90 \%
intensity. 
}
\label{fig:beam2013}
\end{figure}

We estimated the beam size and the main-beam efficiency of the NRO 45-m
telescope based on the observations of SiO masers, and continuum observations of 3C 279 and Saturn in March 2013.
To obtain the beam pattern, we mapped the SiO ($J=1-0$, $v=1$) and
SiO ($J=1-0$, $v=2$) maser emission from an evolved star, R-Leo, in a 
On-The-Fly (OTF) mode. Both $v=1$ and $v=2$ lines are taken simultaneously.
The coordinate of R-Leo is (R.A.[J2000.0], decl.[J2000.0]) = (09:47:33.49, +11:25:43.7).
The parameters for the OTF observations are 
summarized in table \ref{tab:otf}.
We obtained two maps, for each of which we scanned along either the Right Ascension or Declination direction, 
and combined them into a single map in order to reduce the scanning effects.
We adopted a convolution scheme with a Gaussian function to
calculate the intensity at each grid point of the final data 
with a spatial grid size of 10$''$, about a forth of the beam size.
Since the SiO masers observed are very intense, the signal-to-noise
ratios of the velocity-integrated intensity maps of the SiO masers
is extremely high and thus sufficient to inspect the beam pattern
obtained with the new receiver.
In figures \ref{fig:beam2013}a and \ref{fig:beam2013}b, we present the 
velocity-integrated intensity maps of SiO maser emission from 
 R-Leo taken with Z45 
 for both horizontal and vertical polarization components, respectively.
The intensity is normalized to the maximum intensity.
Since the SiO masers come from point sources, the intensity distribution 
should represent the beam pattern at 43 GHz.
We show only $v=1$ transition emission although we take $v=2$ emission
simultaneously. Both data show almost identical beam pattern.
The beam patterns appear to be almost identical to both polarization components.
The SiO intensity distributions are nearly circular above the 
10\% intensity level. However, at lower levels,
there exists asymmetric intensity distribution below the 3 \% level.
This level of the sidelobe is reasonably small and thus we conclude
that the beam pattern is reasonably round.
We note that the beam pattern shows some elongation in the
northern, western, and south-east directions where asymmetric intensity distributions below the 3 \% level
are seen.
We believe that this elongation in the three directions is due to the structure of the sub-reflector stays of the NRO 45-m telescope. 
Similar elongation is observed in the beam patterns obtained with different receivers of the 45-m telescope
(e.g., see Figure 11 of Nakajima et al. 2013a).

The half-power beam width (HPBW) of the telescope is estimated to be
about 37$''$ at the frequency of 43 GHz, by fitting the emission with a two-dimensional
Gaussian function. 
We also measure the beam sizes
 at various elevations summarized in table
\ref{tab:elevation}, and found that the dependence of beam size on
elevation is very small for $32^\circ < {\rm EL} < 62^\circ$.
These measurements were done in 2014 March with images constructed with single scan data.
We note that the beam size was also measured from the continuum observations of
3C 279, and the obtained HPBW is almost equal to that obtained from the SiO maser observations.

\begin{table}
  \caption{Parameters of SiO maser observations}\label{tab:otf}
  \begin{center}
    \begin{tabular}{ll}
     \hline
      Observation mode & OTF \\
      Box size & $3'\times 3'$\\
      Time for scan & 10s\\
      Number of ONs per OFF &  4 \\
      Separation between scans & 5$''$ \\
      Map grid &  10$''$ \\
      OFF-point separation & $-$3$'$ in the R.A. direction. \\
      $T_{\rm sys}$ (K) & 200 \\
      Frequency resolution (kHz) & 30.52 \\
      Convolution function & Gaussian \\
      Observation Time & 20 min \\
      \hline
    \end{tabular}
  \end{center}
\end{table}

 \begin{table}
\begin{center}
  \caption{Beam sizes} \label{tab:elevation}  
    \begin{tabular}{lllllll}
     \hline
      elevation & date and time (JST) & scan & beam size (major)& beam size (minor)& $T_{\rm sys}$ & wind speed (m s$^{-1}$)\\
     \hline
       62$^\circ$.0 & 2014 March 17 20:59 & x & $38.4\pm 0.05$&
		$ 35.6\pm 0.06$ & 110 & 3.6 \\
       57$^\circ$.3 & 2014 March 17  23:33 & x & $38.8 \pm 0.06 $ & $ 35.9\pm
		     0.06$ & 110 & 0.5 \\
       48$^\circ$.5 & 2014 March 18  00:26 & y & $38.4 \pm 0.06$ & $ 36.2 \pm
		     0.06$ & 113 & 0.8 \\
       44$^\circ$.2 & 2014 March 18 00:49 & x & $37.2 \pm 0.06$ & $ 36.2 \pm
		     0.06$ & 116 & 0.3 \\
       39$^\circ$.7 & 2014 March 18 01:13 & y & $37.4 \pm 0.06$ & $ 36.1 \pm
		     0.06$ & 119 & 0.2 \\
       33$^\circ$.8 & 2014 March 18 01:43 & x & $38.4 \pm 0.06 $ & $ 36.4
		     \pm 0.07$ & 125 & 0.3 \\
      \hline
    \end{tabular}
  \end{center}
\end{table}

The main-beam efficiency, $\eta_{\rm mb}$, was calculated to be 72\% at 43 GHz, 
assuming that the brightness distribution
of Saturn to be a uniform disk with an apparent effective radius
of 17$"$.4 and a brightness temperature of 136 K (Ulich 1981).
The observations of Saturn were conducted on June 9 2013.
The effective diameter of Saturn was calculated as
$2 \times ({\rm polar \ radius} \times  
{\rm equatorial \ radius})^{1/2} = 2 \times (9."2 \times 8".3)^{1/2}$. 
The brightness temperature of 136 K was derived by the linear interpolation of the values 
at 31.4 GHz (133.0 K) and 90.0 GHz (149.3 K) to 43 GHz.

We have conducted similar measurements of the main-beam efficiency using Saturn and Mars 
in 2013, 2014, and 2015.  
For all the measurements except observations of Mars done on March 2014, the beam efficiency was measured 
as $0.7-0.75$.  The main-beam efficiency measured on March, 2014 was $\sim 0.6$, smaller by a factor of 1.2. 
A possible cause of this smaller main-beam efficiency might be the effects of snow on February, 2014 when
Nobeyama was hit by terribly heavy snow. But, the actual reason remains uncertain.

\section{Examples of Astronomical Observations}
\label{sec:test}

Here, we show some examples of astronomical observations using the new
receiver Z45.  As for polarization observations, we show their detail 
in a separate paper including the calibration method (Mizuno et al. in preparation).
Since Z45 is a dual-polarization receiver and is connected to 8 sets of
a 4096  channel spectrometer SAM45, one can obtain up to 8 molecular line emission simultaneously.
The rms noise levels can be reduced by a factor of $\sqrt{2}$ to
combine H- and V-polarization components of single lines observed
simultaneously into one.
The typical system temperature of Z45 is less than half of the existing
40 GHz receiver S40 under the same atmospheric conditions. 
Thus, the observation time can be reduced significantly.
To achieve the same noise level, the total observation time using Z45 
is reduced by a factor of about 10 compared to S40.

\subsection{Observations of Various Molecular Lines at 40 GHz}

\begin{figure}
\begin{center}
\includegraphics[width=18cm]{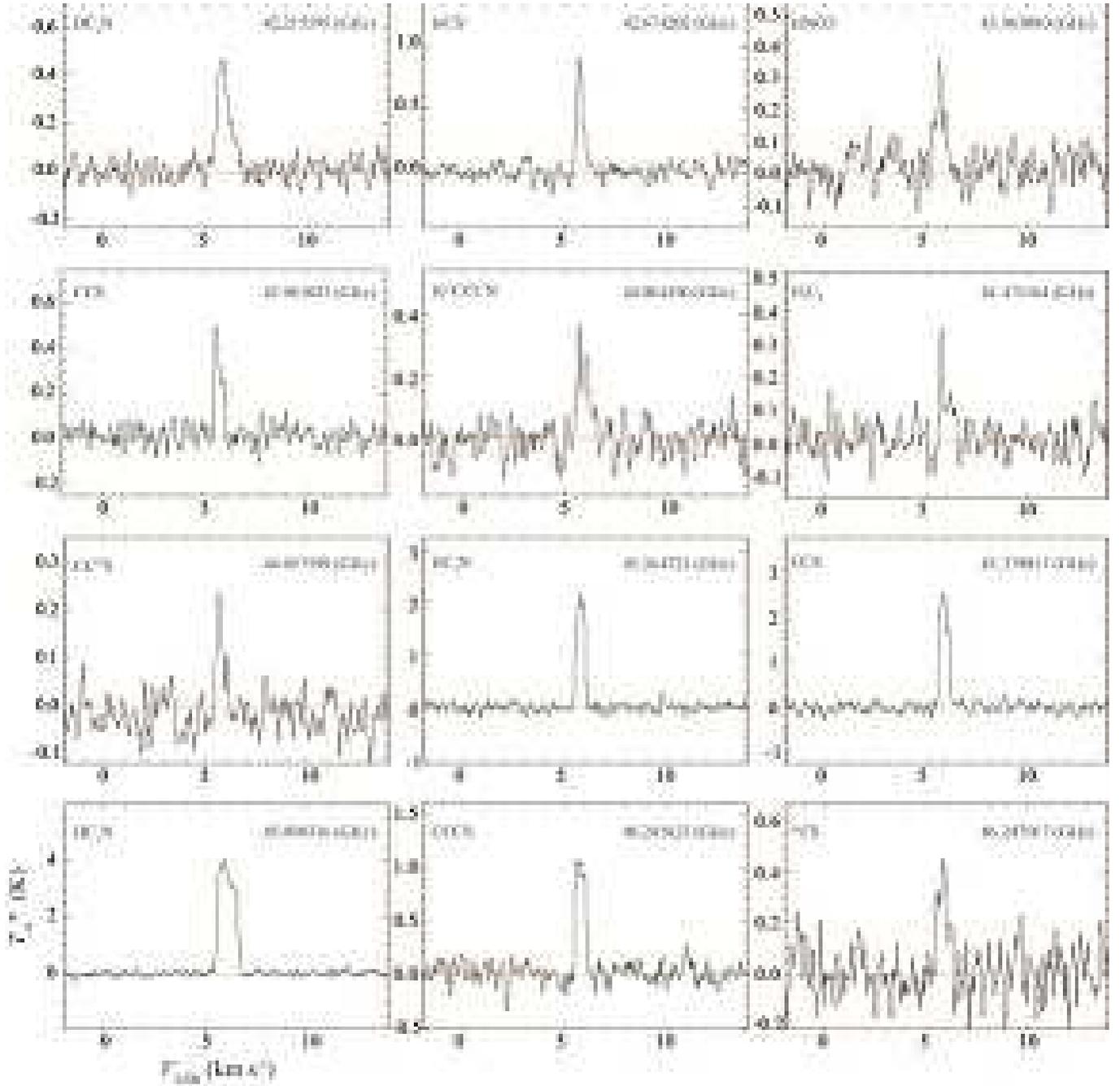} 
\caption{
Twelve molecular lines in table \ref{tab:line} observed toward TMC-1(CP) 
(RA[J2000]=04h41m42.5s and Dec[J2000]=25d41$'$27.0$''$) with the Z45
 receiver. The spectra were integrated for 200-600 seconds. The achieved 
rms noise level was $\Delta T_{\rm rms} =$ 0.03-0.08 K at the  velocity
 resolution $\Delta V\simeq  0.075$ km s$^{-1}$.}
\label{fig:line}
\end{center}
\end{figure}

In figure \ref{fig:line}, we show the results of multi-molecular line 
observations at 40 GHz band toward cyanopolyyne peak of TMC-1 [TMC-1(CP)], 
whose coordinates are 
(R.A. [J2000], decl. [J2000]) = (04:41:42.5, 25:41$'$:27.0$''$).
Frequencies of observed molecular lines are listed in table
\ref{tab:line}, which are adopted from Kaifu et al. (2004).

At the 40 GHz band, CS ($J=1-0$) is one of the important molecular lines
to  especially trace the dense molecular gas. However, 
CS ($J=1-0$) and C$^{34}$S ($J=1-0$) at frequencies of 48.990978 GHz
and 48.206915 GHz are out of bandwidth and cannot be observed using Z45
for now.  $^{13}$CS, isotopologue of CS, at 46.247567GHz can be
observed.
We plan to extend the bandwidth to cover up to $\sim 50$ GHz
in future.

Observations were done in 2014 March and April. As a back end we used SAM45 with frequency
resolution of 3.81 kHz, corresponding to 0.025 km s$^{-1}$ velocity
resolution
at 45 GHz. The typical system temperature during the observations
was 130$-$160 K.
We use the position offsetting ($\Delta R.A., \Delta decl.$) 
=(30$'$, 30$'$) from TMC-1 (CP) as a emission free off position.
We obtained CCS, HC$_3$N, and HC$_5$N simultaneously in 2014 March. 
Other lines were also  obtained  simultaneously in 2014 April. 
The telescope pointing was checked by observing
a SiO maser source, NML Tau, and the typical pointing offset
were within  $5''$ during the whole observation period.
The standard chopper wheel method was used to
convert the output signal into the antenna temperatures ($T_A^*$)
and corrected for the atmospheric attenuation.
The integration time of the source is about 200$-$600 seconds.
The achieved rms noise levels are in the range from 0.03 to 0.08 K
at the velocity resolution of about 0.075 km s$^{-1}$.

\begin{table}
\caption{Molecuar lines observed toward TMC-1.}\label{tab:line}
\begin{center}
\begin{tabular}{lll}
\hline
Molecule & Transition & Rest Frequency (GHz)\\
\hline
DC$_{3}$N&$J=5-4$& 42.215595 \\
HCS$^{+}$&$J=1-0$& 42.674202 \\
HNCO& $J_{KaKc}=2_{02}-1_{01}$ $F=3-2$ \& $F=2-1$ &43.963000\\
CCS&$J_{N}$=3$_{4}-2_{3}$& 43.981023 \\
H$^{13}$CCCN&$J=5-4$	& 44.084190 \\
H$_{2}$C$_{4}$&$J_{KaKc}=5_{15}-4_{14}$& 44.471164 \\
CC$^{34}$S&$J_{N}$=4$_{3}-3_{2}$ & 44.497599 \\
HC$_{5}$N&$J=17-16$& 45.264721 \\
CCS&$J_{N}$=4$_{3}-3_{2}$&45.379033 \\
HC$_{3}$N&$J=5-4$&45.490316 \\
CCCS&$J=8-7$&46.245623\\
$^{13}$CS&$J=1-0$&46.247567\\
\hline
\end{tabular}
\end{center}
\end{table}

In figure \ref{fig:line}, we show the obtained line profiles of 12 molecular
lines toward TMC-1 (CP). Carbon-chain molecules such as CCS, CCCS, HC$_3$N,
and HC$_5$N show strong emission. These molecules are known to be
abundant in the early stage of prestellar evolution.
Because of high sensitivity of Z45, we
can also  detect weak lines such as CC$^{34}$S ($J_N=4_3-3_2$), DC$_3$N, and
H$^{13}$CCCN. 
Using these lines, we can estimate the optical depth (e.g., CCS and
CC$^{34}$S) and [D/H] to understand deuterium fractionation 
in molecular clouds (e.g., HC$_3$N and DC$_3$N). 
We note that the CC$^{34}$S, CCCS, and HCS$^+$ peak intensities shown in figure \ref{fig:line} are
about 40, 50, and 100 \% stronger than those of Kaifu et al. (2000), respectively.
We suspect that this is partly due to the flux calibration problem of the back end system SAM45 (e.g., non-linearity).
Similar problems were reported for other observations using with the SAM 45 backend system (e.g., Nakamura et al. 2014).
Therefore, the line intensities shown here may have significant errors.
For the Zeeman measurements, we use the newly-developed spectrometer, PolariS, 
instead of SAM45.

\subsection{Mapping Observations}

\begin{figure}
\begin{center}
\includegraphics[width=15cm]{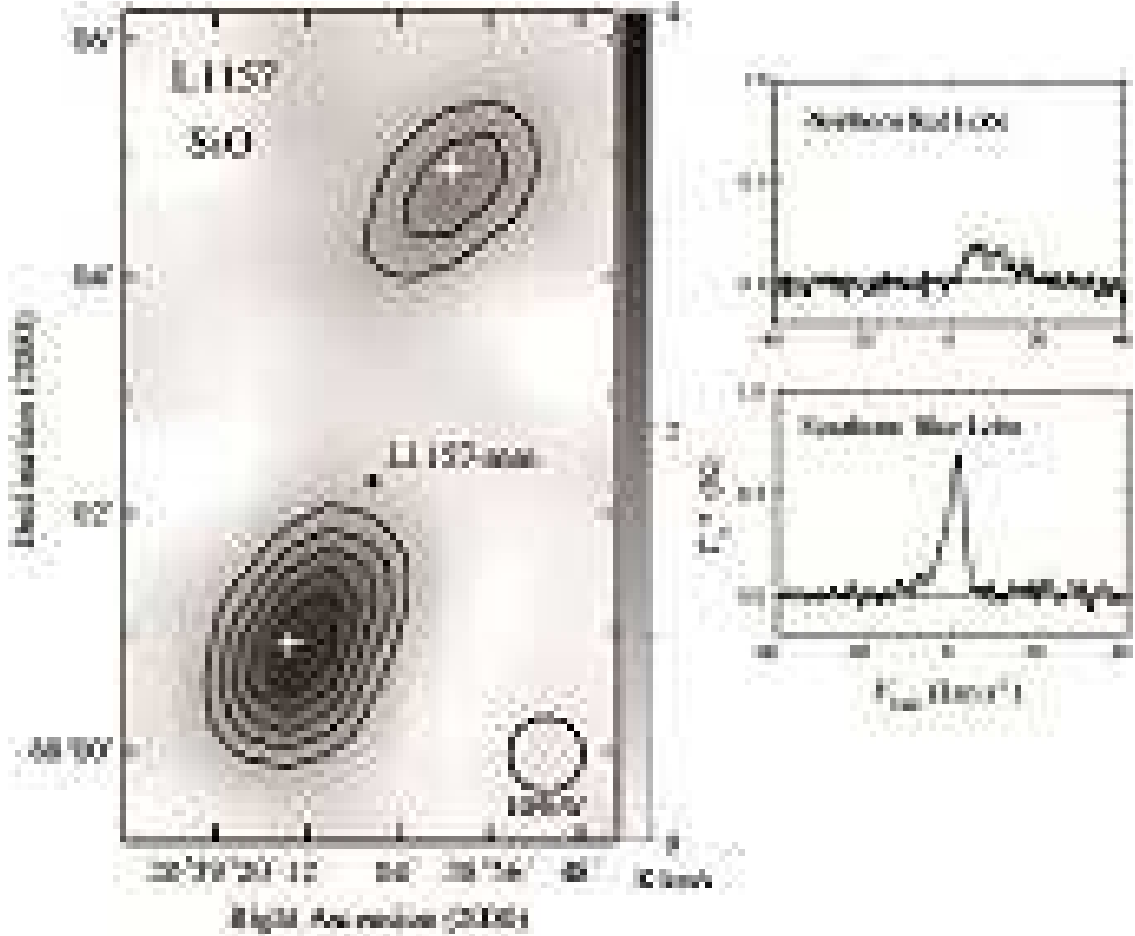} 
\caption{
Intensity map of the SiO($J=1-0,v=0$) emission line at 43.4 GHz
around the outflow source L1157 integrated over the velocity 
range -10 $<$ V$_{\rm LSR}$ $<$ 20 km s$^{-1}$.
Filled circle denotes the position of the driving source 
L1157-mm (RA2000=20h39m06.19s, Dec2000=68d02'15.9$''$, Umemoto et al. 1992, Bachiller et al. 2001).
Contours start from $\int T_A^*dV$ = 1.0 K km s$^{-1}$ with an increment of 0.5
 K km s$^{-1}$. HPBW of Z45 is shown in cycle at the right bottom corner.
Two condensations in the north and south to the driving source
correspond to the outflow lobes (Umemoto et al. 1992, Mikami 1992). 
SiO spectra observed in the blue lobe (sampled at 20h39m13.6s and 68$^\circ$00$'$54$''$) 
and red lobe (20h38m59.3s and 68$^\circ$04$'$54$''$) are shown in the
 right panels.
Observed points are also indicated by the crosses.
}
\label{fig:L1157}
\end{center}
\end{figure}

Figure \ref{fig:L1157} is an example of mapping observations toward
L1157 in SiO ($J=1-0, v=0$).
We carried out the observations in 2014 April in the OTF mode.
The coordinate of L1157 is (R.A.[J2000.0], decl.[J2000.0])=
(20:39:06.19, 68d2$'$15.9$''$).
We spent about 3 hours to obtain the map. 
We used SAM45 at frequency resolution of 3.81 kHz as a back end. 
The typical system temperature during the observations was 120$-$140 K.
We use the position offset ($\Delta$R.A., $\Delta $decl.) =(30$'$, 30$'$)
from the L1157 position as a emission free off position.
The telescope pointing was checked every 1 hr by observing
a SiO maser source, IRC+60427, and the typical pointing offset
was within $5''$ during the whole observation period.
The standard chopper wheel method was used to
convert the output signal into the antenna temperatures ($T_A^*$)
and corrected for the atmospheric attenuation.

L1157 is the Class 0 object which is associated with a powerful outflow
(Umemoto et al. 1992, Bachiller et al. 2001).
SiO($J=1-0,v=0$) line can be often used to detect young outflow.
Our map clearly shows the bipolar outflow  from L1157-mm. 
The extent of the outflow lobes looks similar to the maps of higher
transition SiO lines presented in Fig. 7 of Bachiller et al.(2001). 
We can also detect the faint redshifted lobe in northern part of
L1157-mm.

Another example of mapping observations is presented in figure 1(b) and 1(c)
of Nakamura et al. (2014), 
who showed the CCS ($J_N=4_3-3_2$) and HC$_3$N ($J=5-4$) integrated
intensity maps of about $10'\times 20'$ area of Serpens South infrared
dark cloud
(see figure \ref{fig:serps} for CCS integrated intensity map).
Nakamura et al. (2014) showed that pre-protostellar clumps in Serpens South
has very strong CCS and HC$_3$N emission.
It confirms that these carbon-chain molecules are abundant in prestellar
phase and suitable for searching for future star formation sites in
molecular clouds. 
High sensitivity of Z45 allows us to carry out these large-scale mapping
observations of nearby star-forming regions in reasonable amount of
observational time.

\section{Conclusions}
\label{sec:con}

We developed a dual-polarization HEMT receiver system in
the 40-GHz band and installed in the NRO 45-m telescope.
We named the new receiver Z45, which is designed to conduct 
Zeeman measurements of CCS ($J_N=4_3-3_2$).
The receiver system is designed to conduct polarization
 observations by taking the cross correlation of two linearly-polarized components, 
from which we can process full-Stokes spectroscopy.
A linear-polarization receiver system has a smaller
contribution of instrumental polarization components to the Stokes
$V$ spectra than that of the circular polarization system, so that 
our system has an advantage of measuring the line-of-sight magnetic field strength,
using the Stokes $V$ spectra.
Table \ref{tab:summary} summarizes the characteristics of Z45.
The typical receiver noise temperature including atmosphere is in the range from 100 K to 150 K for the
frequency of 42-46 GHz.
The receiver has an intermediate frequency (IF)
 band of 4.0$-$8.0 GHz. 
We developed ortho-mode transduser and corrugated horn.
The receiver system is connected to two back ends, SAM45 and PolariS.
SAM45 can be used for standard observations like a position-switch (PS) and
 on-the-fly (OTF) mapping observations.
Zeeman measurement can be done with the PolariS spectrometer. On the basis
 of SiO maser observations and continuum observations of Saturn,
we also measured beam size and main beam efficiency to be 37$''$ and 0.72
 in the 43 GHz. To demonstrate the performance of the new system, we
 have carried out several observations such as PS and OTF observations,
which demonstrate the  high sensitivity of Z45 allows us to map over
 wider field and detect faint lines in a short time.

\begin{table}
\begin{center}
  \caption{Characteristics of Z45} \label{tab:summary}  
    \begin{tabular}{lll}
     \hline
 Type of receiver & HEMT \\
 RF frequency & 42$-$ 46 GHz\\
 IF frequency & 4$-$ 8 GHz \\
 $T_{\rm RX}$ & $\sim$ 50 K \\
 Polarization & dual linear \\
 back end & SAM45, PolariS \\
 Beam size (HPBW) & 37$"$ \\
 Beam efficiency & 0.72 (43GHz) \\
      \hline
    \end{tabular}
  \end{center}
\end{table}

This work is supported in part by a Grant-in-Aid for Scientific 
Research of Japan (24244017) and the National Science Foundation under
Grant (NSF PHY11-25915).
We are grateful to the staffs at the Nobeyama Radio Observatory 
for operating the NRO 45-m telescope. 
The Nobeyama Radio Observatory is a branch of the National
Astronomical Observatory of Japan, National Institutes of Natural Sciences;
(\verb/http://www.nro.nao.ac.jp/index-e.html).

\clearpage


%
%
%
%

\end{document}